\documentclass[%
 aip,
 sd,%
 amsmath,amssymb,
twocolumn,
10pt,
a4paper,
nofootinbib,
tightenlines
]{revtex4-1}

\usepackage{
    amsmath,
    amssymb,
    amsthm,
    bm,
    graphicx,
    caption,
    enumerate,
    xcolor,
    mathtools,
    dsfont,
    bm,
    listings,
    float,
    stackengine,
    makecell,
    booktabs,
    dcolumn,
    array,
    bbm
}

\usepackage{tabularx}
\usepackage{array}

\usepackage[compatible]{algpseudocode}
\usepackage[section]{algorithm}
\usepackage[disable]{todonotes}
\usepackage[percent]{overpic}

\usepackage[utf8]{inputenc}
\usepackage[pdffitwindow=false,
            plainpages=false,
            pdfpagelabels=true,
            pdfpagemode=UseOutlines,
            pdfpagelayout=SinglePage,
            bookmarks=false,
            colorlinks=true,
            hyperfootnotes=false,
            linkcolor=blue,
            urlcolor=blue!30!black,
            citecolor=green!50!black]{hyperref}

\makeatletter
\newcommand{\multiline}[1]{%
  \begin{tabularx}{\dimexpr.9\columnwidth-\ALG@thistlm}[t]{@{}X@{}}
    #1
  \end{tabularx}
}
\makeatother

\makeatletter
\newcommand\fs@nobottomruled{\def\@fs@cfont{\bfseries}\let\@fs@capt\floatc@ruled
  \def\@fs@pre{\hrule height.8pt depth0pt \kern2pt}%
  \def\@fs@post{}
  \def\@fs@mid{\kern2pt\hrule\kern2pt}%
  \let\@fs@iftopcapt\iftrue}
\makeatother      

\makeatletter
\newcommand\fs@notopruled{\def\@fs@cfont{\bfseries}\let\@fs@capt\floatc@ruled
  \def\@fs@pre{}%
  \def\@fs@post{\kern2pt\hrule\kern2pt}
  \def\@fs@mid{}%
  \let\@fs@iftopcapt\iftrue}
\makeatother   

\newcommand{\R}{\mathbb{R}}                                     
\newcommand{\X}{\mathbb{X}}                                     
 
\DeclareMathOperator*{\argmax}{arg\,max}

\mathtoolsset{centercolon} 
\newtheorem{theorem}{Theorem}[section]

\theoremstyle{definition}
\newtheorem{example}[theorem]{Example}

\newtheorem*{remark*}{Remark}
\newtheorem*{example*}{Example}
\newtheorem*{summary*}{Summary}

\newcolumntype{L}[1]{>{\raggedright\let\newline\\\arraybackslash\hspace{0pt}}m{#1}}
\newcolumntype{C}[1]{>{\centering\let\newline\\\arraybackslash\hspace{0pt}}m{#1}}
\newcolumntype{R}[1]{>{\raggedleft\let\newline\\\arraybackslash\hspace{0pt}}m{#1}}
\newcolumntype{H}{>{\setbox0=\hbox\bgroup}c<{\egroup}@{}}

\definecolor{boxback}{gray}{0.95}

\title{Data-driven Computation of Molecular Reaction Coordinates
}

\begin{document}

\title[Data-driven Computation of Molecular Reaction Coordinates]{Data-driven Computation of Molecular Reaction Coordinates}

\author{Andreas Bittracher}
\email[]{bittracher@mi.fu-berlin.de}
\affiliation{ 
Department of Mathematics, Freie Universit\"at Berlin, 14195 Berlin, Germany
}
\author{Ralf Banisch}%
\email[]{ralf.banisch@fu-berlin.de}
\affiliation{ 
Department of Mathematics, Freie Universit\"at Berlin, 14195 Berlin, Germany
}
\author{Christof Sch\"utte}
\email[]{christof.schuette@fu-berlin.de}
\affiliation{ 
Department of Mathematics, Freie Universit\"at Berlin, 14195 Berlin, Germany
}
\affiliation{ 
Zuse Institute Berlin (ZIB), 14195 Berlin, Germany
}%

\begin{abstract}

The identification of meaningful reaction coordinates plays a key role in the study of complex molecular systems whose essential dynamics is characterized by rare or slow transition events. 
In a recent publication, precise defining characteristics of such reaction coordinates were identified and linked to the existence of a so-called transition manifold. This theory gives rise to a novel numerical method for the pointwise computation of reaction coordinates that relies on short parallel MD simulations only, but yields accurate approximation of the long time behavior of the system under consideration.

This article presents an extension of the method towards practical applicability in computational chemistry. It links the newly defined reaction coordinates to concepts from transition path theory and Markov state model building. The main result is an alternative computational scheme that allows for a global computation of reaction coordinates based on commonly available types of simulation data, such as single long molecular trajectories, or the push-forward of arbitrary canonically-distributed point clouds. It is based on a Galerkin approximation of the transition manifold reaction coordinates, that can be tuned to individual requirements by the choice of the Galerkin ansatz functions.
Moreover, we propose a ready-to-implement variant of the new scheme, that computes data-fitted, mesh-free ansatz functions directly from the available simulation data. The efficacy of the new method is demonstrated on a small protein system.

\end{abstract}

\maketitle

\section*{Introduction}

In recent years, it has become possible to numerically explore the chemically relevant slow transition processes in systems with several thousands of atoms.
This was made possible due to the increase of raw computational power and deployment of specialized computing architectures \cite{ShawEtAl}, as well as by the development of accelerated integration schemes that bias the dynamics in the favor for the slow transition processes, yet preserve the original statistics \cite{Aristoff2014,Elber,metadynamics-review}.

To obtain chemical insight about the essential dynamics of the system, this vast amount of high-dimensional data has to be adequately processed and filtered.
One desirable goal often is a simplified model of the mechanism of action, in which the fast, unimportant processes are averaged out or otherwise disregarded.
One way is to construct \emph{kinetic} models of the system, i.e., identifying metastable reactant-, product- and possibly intermediate states, and reducing the dynamics to a jump process between them. Under certain regularity assumptions on the root model that are readily fulfilled, such a model can be built in an automated, data-driven fashion \cite{A19-1,SchSa13}.
However, the simplicity of the resulting so-called \emph{Markov state model} (MSM) comes with a price: since the long-time relaxation kinetics is described just by jumps between finitely-many discrete states, any information about the transition process and its dynamical features is lost.

An alternative collection of approaches, to which this paper ultimately contributes, thus aims at the automated identification of good \emph{reaction coordinates} or \emph{order parameters}, mappings from the full to some lower-dimensional, but still \emph{continuous} state space, onto which the full dynamics can be projected without loss of the essential processes. Often enough, this reaction coordinate alone (i.e., without the corresponding dynamical model) already contains more valuable chemical information than the kinetic models, as for example the free energy profile along the reaction coordinate allows the determination of the activation energy of the respective transition process \cite{RC_overview}.

The systematic and mathematically rigorously motivated construction of reaction coordinates is an area of active research, for an overview see Ref.~\onlinecite{RC_overview}.
Where it is available, chemical expert knowledge can be used to guide the construction \cite{CaTh93,SoEtAl96}.
In the context of \emph{transition path theory} (TPT) \cite{VE06,tpt2010}, the committor function is known to be an ideal reaction coordinate \cite{LuVE14} for transitions between preselected metastable sets.
Related to this, approximations to the dominant eigenfunctions of the transfer operator are also often considered ideal reaction coordinates \cite{SchSa13,CN14,tika2013}, which has been confirmed in Ref.~\onlinecite{FGH14a} for a subclass timescale separated systems.
However, the computation of both committor functions and transfer operator eigenfunctions is infeasible for very high-dimensional systems. Moreover, the authors have recently shown that said eigenfunctions yield redundant reaction coordinates, in the sense that often a further reduction is possible \cite{Bittracher2017}.

In the same work, the authors identified necessary characteristics that reaction coordinates have to exhibit in order to retain the slow processes (a ``quality criterion''). In short, it must be possible to relate them to the dominant transfer operator eigenfunctions in a specific non-linear way. 
However, as we will see, the criterion is also interpretable in the context of TPT.

What is more, it was shown that the existence of reaction coordinates that fulfill the quality criterion is tied to the existence of a so-called \emph{transition manifold} $\mathbb{M}$, a low-dimensional manifold in the function space $L^1$. The property that defines $\mathbb{M}$ is that, on moderate time scales $t_\text{fast}< t\ll t_\text{slow}$, the \emph{transition density functions} of the dynamics concentrate around~$\mathbb{M}$. A firm mathematical theory for the existence and identification of reaction coordinates was developed around this transition manifold.

The main practical result of Ref.~\onlinecite{Bittracher2017} was the insight that any parametrization of $\mathbb{M}$ can be turned into a good reaction coordinate. A numerical algorithm was proposed that allows the pointwise computation of this reaction coordinate and only requires the ability to generate trajectories of the aforementioned moderate length that start at the desired evaluation point.

While the method has a solid theoretical foundation and is directly applicable in many cases, there yet exists a certain gap between the theoretical advantages and the practical applications of the proposed scheme:
While the ability to efficiently compute the reaction coordinate only in specific points is quite remarkable, in practice one often wishes to learn the reaction coordinate in \emph{all} of the accessible state space (i.e., where pre-generated simulation data is available), as the location of the ``interesting'' points is unknown in advance.
The originally proposed method cannot compute the reaction coordinate from dynamical ``bulk data'' -- such as long equilibrated trajectories or the push-forward of point clouds that sample the canonical ensemble -- that is preferably generated by contemporary simulation methods and software.

In the present work we attempt to close this gap by proposing an alternative, purely data-driven algorithm for computing the transition manifold reaction coordinate. It is based on a classical Galerkin approximation of the reaction coordinate with freely selectable ansatz space. Its numerical realization requires only a so-called transition matrix between its discretization elements. A wide variety of techniques for building MSMs and similar algorithms is available for construction of this matrix from the aforementioned types of bulk data \cite{MSMBuilder,PriEtAl11,tika2013}.
This makes it possible to transfer many techniques from the extensive toolbox of MSMs, as for example the use of customized Galerkin ansatz spaces explicitly adapted for molecular dynamical problems \cite{ViNoKe15}.
Further, this makes our approach instantly applicable whenever the construction of an (arguably less informative) MSM is possible.

Finally, with the objective to create an algorithm that requires only a minimum of a priori information about the system, we propose a very practical implementation of this Galerkin approximation that constructs a mesh-free set of Voronoi cell-based ansatz functions directly from the available simulation data.
Interestingly, the task of optimally choosing the Voronoi centers leads to two well-known and highly scalable algorithms from data mining, namely the k-means clustering algorithm and Poisson disk sampling algorithm, depending on the chosen error measure.
We demonstrate the efficacy of this method by identifying chemically interpretable essential degrees of freedom of a 66-dimensional model of alanine dipeptide, and a 1600-dimensional model of the fast-folding protein NTL9.

The paper is organized as follows: In Section \ref{sec:basics} the basic concepts of timescale-separated systems and reaction coordinates are introduced. Also, our central quality criteria for the characterization of good reaction coordinates is derived and a comparison with TPT is drawn.
 Section \ref{sec:methods} introduces the concept of transition manifolds and explains the local burst-based algorithm. In Section \ref{sec:Galerkin}, the new Galerkin approximation of the transition manifold reaction coordinate is derived as well as the Voronoi-based implementation. Section \ref{sec:examples} demonstrates the application of our new method to a simple synthetic example system, as well as to the realistic molecular systems.
Concluding remarks and an outlook can be found in \ref{sec:conclusions}.


\section{Characterization of good reaction coordinates}
\label{sec:basics}

%
%
%
%
%
%
%
%

\subsection{Metastable molecular dynamics}

We model our molecular dynamical system as a continuous-time stochastic process $\bm{X}_t$ on some high-dimensional state space $\X\subset\R^n$. Here $\X$ may consist of either full Cartesian atomic coordinates or some other suitable degrees of freedom that adequately describe the micro state of the system. We require the process to fulfill common technical assumptions from the Markov approach to molecular dynamics\cite{PriEtAl11,SchueHuiDeu01}, namely Markovianity, ergodicity and time-reversibility. Aside from that, the specific dynamical law that governs the evolution of $\bm{X}_t$ is ultimately arbitrary, but we in general think of $\bm{X}_t$ as a ``random walk in a potential energy landscape''. 
The first example that comes to mind would be the Smoluchowski dynamics (also called \emph{overdamped Langevin dynamics})
\begin{equation}
\label{eq:SmolDyn}
d\bm{X}_t = -\nabla V(\bm{X}_t)dt + \sqrt{2\beta^{-1}} \bm{W}_t~,
\end{equation}
where $V$ denotes the potential energy function, $\beta=1/k_BT$ the inverse temperature, and $\bm{W}_t$ a standard Wiener process. However, our theory can also be applied to the non-overdamped Langevin dynamics (projected onto the positional degrees of freedom), or any other thermostated molecular dynamics that samples the stationary probability density 
$$
\rho(x) = Z^{-1}e^{-\beta V(x)}~.
$$
Here, $Z=\int_\X e^{-\beta v}$ is a normalizing constant.

\subsection{Reaction coordinates}
Formally, a reaction coordinate is a low-dimensional variable of the full system, i.e. a smooth function $\xi:\R^n\rightarrow\R^k$ with $k \ll n$. In practice, $k$ will often be only one- or two-dimensional, and correspond to some chemically interpretable quantity, e.g. a certain collection of backbone dihedral angles in a peptide, or the distance between important functional groups. The \emph{reduced} or \emph{projected system} is then given by $\xi(\bm{X}_t)$, which is now a stochastic process on $\R^k$.

While the map $\bm{x}\mapsto \xi(\bm{x})$ describes the pointwise projection of the system, the projection of densities that evolve with the system is described by the Zwanzig projection operator\cite{Zwa61}, denoted by $\mathcal{Q}_\xi$:
$$
\big(\mathcal{Q}_\xi p\big) (z) = \frac{1}{W(z)} \int_\X p(x)\rho(x) \delta_z\big(\xi(x)\big)~dx~,
$$
where $\delta_z$ is the delta distribution and $W(z)$ is a normalization term.
Its action can be described as follows: if for some time $t$ the random variable $\bm{X}_t$ is distributed according to some density $p_t$, then the random variable $\xi(\bm{X}_t)$ is distributed according to $\mathcal{Q}_\xi p_t$, which is a density over $\R^k$.

By the definition above, any function over $\R^n$ may be called a reaction coordinate. Thus, one of the key questions we aim to answer in this article is: what criterion distinguishes ``good'' from ``bad'' reaction coordinates?

In many MD systems, the chemically interesting reaction processes correspond to transitions between two or more \emph{metastable states}, regions of state space that ``trap'' the dynamics for long times before a sudden transition to another metastable state occurs. Typical examples include protein- and peptide folding, receptor-ligand binding and conformational change of large biomolecules. 
It is customary to picture these transitions as occuring along certain \emph{transition pathways} in the potential energy landscape, but there is no uniformly accepted definition of these pathways. Proposed variants include the minimum energy path \cite{Fuk70}, minimum free energy path \cite{MaFiVaCi06} and the principal curve \cite{VaVe09}.
By a first intuitive definition, reaction coordinates should thus describe the ``progress of the reaction along the transition pathway''.
A common computational scheme thus goes as follows:
\begin{enumerate}
\item Compute the transition pathway (using for example the string method\cite{E2002}).
\item Parametrize the transition pathway.
\item Project the state space onto the transition pathway.
\end{enumerate} 
The value of the parametrization in a projected point then gives the reaction coordinate value in that point.

However, due to the ambiguous concept of transition pathways, this approach lacks rigor. Variants of transition pathways that are based on local features of the energy landscape only, such as the minimum energy pathway, can be shown to fail to describe the (global) slow transition processes\cite{VaVe09}.
Moreover and most importantly, the question of how to globally project state space onto the transition path in a ``dynamically correct'' way remains unanswered. A nearest-point projection, as is for example used in the definition of principal curves, can be shown to fail with simple example systems.

Thus, in order to find a rigorous criterion for good reaction coordinates, we need to take a closer look at the ``global'' stochastic evolution of $\bm{X}_t$ and its slow parts. We will however eventually come back to the picture of potential energy surfaces and interpret our criterion with regard to transition pathways (see Example \ref{ex:BananaPot}).

\subsection{The transfer operator}
\label{sec:transferoperator}

Regardless of the specific dynamical model, the stochastic evolution of $\bm{X}_t$ is entirely described by its transition probability density $p^t$: Given a starting point $x$, the probability density for finding the system at some point $y$ after time $t\geq 0$ is denoted by $p^t(x,y)$:
$$
\bm{X}_0=x\quad\Rightarrow\quad \bm{X}_t\sim p^t(x,\cdot)~,
$$
where ``$\sim$'' means ``distributed according to''.
$p^t(x,\cdot)$ can be estimated by starting a large number of parallel simulations of the stochastic dynamics, all with starting point $x$, and estimating the resulting end point density (for example using histogram or kernel density estimation methods).

With $p^t$, the evolution of a general starting density $X_0\sim u_0$ can then be expressed as
\begin{equation}\label{eq:transferoperator}
u_t(x) = \int_\X u_0(y) p^t(y,x)~dy =: \mathcal{P}^t u_0(x)~.
\end{equation}
The operator $\mathcal{P}^t$ is known as the \emph{Perron-Frobenius operator} or \emph{transfer operator} of the system. In the case of the Smoluchowski dynamics \eqref{eq:SmolDyn}, it is equal to the solution operator of the associated Fokker-Planck equation.

We see that $p^t$ and by extension $\mathcal{P}^t$ describe the complete stochastic evolution.
While the analytical derivation of $p^t_x$ and $\mathcal{P}^t$ is possible only for the most simple of systems (for example for the Ornstein-Uhlenbeck process\cite{KNKWKSN17}), they will play the central role in the description of slow sub-processes of the dynamics and the computation optimal reaction coordinates. 

\begin{remark*}
Closely related to $\mathcal{P}^t$ is the \emph{Koopman operator}, defined by
\begin{equation}
\label{eq:koopmanoperator}
\mathcal{K}^t \eta_0(x) = \int_\X \eta_0(y) p^t(x,y)~dy~.
\end{equation}
It acts as the push-forward of observables $\eta_0:\X\rightarrow \mathbb{R}$ under $\bm{X}_t$, i.e. is the conditional expectation of $\eta_0(\bm{X}_t)$, provided we started in $x$ at time $t=0$:
$$
\eta_t(x) = \mathcal{K}^t \eta_0(x) = \mathbb{E}\big[ \eta_0(\bm{X}_t)~|~X_0=x\big]~.
$$
This operator will be of relevance later when we describe the numerical computation of reaction coordinates. 
\end{remark*}

\subsection{Dominant timescales}

Under fairly general conditions, it can be shown that the spectrum of $\mathcal{P}^t$ consists of discrete real eigenvalues 
$$
1 = \lambda_0 < \lambda_1^t \leq \lambda_2^t\leq\cdots~,
$$
and that the eigenvalue $\lambda_0=1$ is simple and belongs to the eigenfunction $\rho$ (the stationary density). We denote by $v_i$ the eigenfunction belonging to $\lambda_i^t$. Except $\lambda_0$, all eigenvalues decay exponentially as $t\rightarrow \infty$, which corresponds to the relaxation of the process towards the stationary ensemble, regardless of the starting density. The relaxation rate of the $i$-th slowest process, known as the \emph{$i$-th implied timescale}\cite{PriEtAl11}, is given by
\begin{equation}\label{eq:ImpliedTimescales}
t_i = -1/\log\big(\lambda_i^t\big)~.
\end{equation}

We from now on assume the system to possess $d$ slow sub-processes, typically (but not necessarily) corresponding to the rare transitions between $d$ metastable sets, and that we are primarily interested in accurately describing these slow processes.
In this case, the dominant $d+1$ eigenvalues $\{\lambda_0^t,\ldots,\lambda_d^t\}$ will be positive and separated from the remaining eigenvalues by a \emph{spectral gap}, i.e. $\lambda_d^t \gg \lambda_{d+1}^t$.
We can then express the action of the operator $\mathcal{P}^t$, and thus the stochastic evolution of the process, in terms of the dominant eigenfunctions:
\begin{align*}
\mathcal{P}^tu_0 \approx \sum_{i=1}^{d}\lambda_i^t c_i v_i~,
\end{align*}
where $c_i= \int u_0 v_i$.
This means that the information about the long-term evolution of the slow processes is entirely contained in the $d$ dominant eigenpairs $(\lambda_i^t,v_i)$. 
Consequently, we consider the preservation of the dominant eigenpairs under projection onto the reaction coordinate a suitable objective for optimally choosing the reaction coordinate.

\begin{remark*}
The dominant eigenpairs of the transfer operator are also the primary object of interest in the Markov approach to coarse graining molecular dynamics, as mentioned in the introduction. Here, the goal is to use the eigenfunctions to build a discrete \emph{Markov State Model} (MSM) \cite{A19-1,SchSa13,pande2010everything}, which replaces the original molecular dynamics by a finite-state Markov jump process between the metastable states. 
Though all information about the transition regions and -paths is lost by this approach, the long-time transition rates between the states are preserved. These models have been successfully applied to a wide range of real-life molecular systems \cite{Schuette1999,pande2010everything,msm_milestoning,CN14}.

The reaction coordinate we will ultimately define and compute will preserve the dominant eigenfunctions, thus the projected process $\bm{\xi}(\bm{X}_t)$ also still contains all the information about the long-term transition processes. 
In this sense, the motivation of ours and the MSM approach are deeply linked.
\end{remark*}

\subsection{A criterion for good reaction coordinates}
\label{sec:criterion}

Our investigation so far points out an apparent discrepancy in the concurrent understanding of what criterion defines ``good'' reaction coordinates:
On one hand, it is a common perception that good reaction coordinates should parametrize some sort of transition pathway, along which a reaction event progresses ``most likely''.
On the other hand, if one is interested in the longterm behaviour of the system, the projection onto the reaction coordinate must preserve the slowest processes, so a definition based on the dominant eigenpairs of the transfer operator seems natural. However, this second requirement appliable to very general and not necessarily metastable systems, thus does not even require the existence of a transition pathway in the classical sense.

We will now see that these two viewpoints can still be unified, and that there exists a criterion for good reaction coordinates based on the transfer operator that also leads to the parametrization of the transition pathway.

Let the \emph{projected transfer operator}, transporting probability densities of the projected process~$\xi(\bm{X}_t)$, be denoted by $\mathcal{P}_\xi^t$. Let $(\mu_i^t,w_i)$ denote the eigenpairs of $\mathcal{P}_\xi^t$.
By the preceeding reasoning, we now call $\xi$ a \emph{good reaction coordinate}, if for the dominant eigenpairs holds
\begin{alignat}{3}
\mu_i^t &\approx \lambda_i^t~,\quad &&i=0\ldots,d \label{eq:optcrit1} \\
\intertext{i.e. the full and projected dominant eigenvalues are similar, and}
v_i(\cdot) &\approx w_i \big( \xi(\cdot)\big)~,\quad &&i=0\ldots,d~ \label{eq:optcrit2}\tag{CI},
\end{alignat}
i.e. the eigenfunctions of $\mathcal{P}^t$ can be approximately reconstructed from the eigenfunctions of $\mathcal{P}^t_\xi$ and $\xi$. This way, all information about the $d$ slowest processes is contained in $\xi(\bm{X}_t)$.

It has been shown in Ref.~\onlinecite{Bittracher2017} that \eqref{eq:optcrit1} follows from \eqref{eq:optcrit2}, so \eqref{eq:optcrit2} is a sufficient criterion for good reaction coordinates (in the sense of preserving the long timescales). If the approximation in \eqref{eq:optcrit2} holds sharp, we call $\xi$ an \emph{optimal reaction coordinate}.

\begin{figure}
\centering

\begin{tabular}{c c}
\bf{(a)} & \bf{(b)}\\
\includegraphics[scale=1]{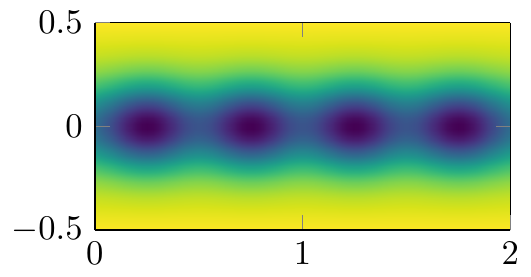} & 
\raisebox{-3mm}{\includegraphics[scale=1]{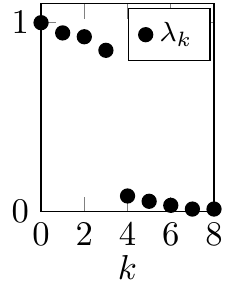}}
\end{tabular}

\begin{tabular}{c c}
\bf{(c)} &
\begin{minipage}{.48\textwidth}
\begin{minipage}{.48\textwidth}
\includegraphics[scale=.9]{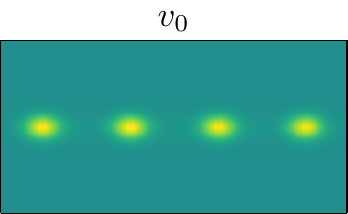}
\end{minipage}
\begin{minipage}{.48\textwidth}
\includegraphics[scale=.9]{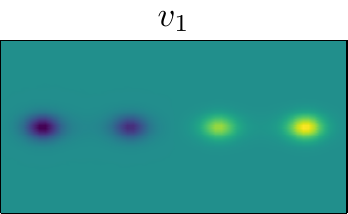}
\end{minipage}

\vspace{0.8em}
\begin{minipage}{.48\textwidth}
\includegraphics[scale=.9]{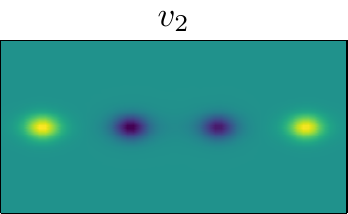}
\end{minipage}
\begin{minipage}{.48\textwidth}
\includegraphics[scale=.9]{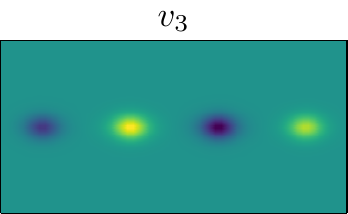}
\end{minipage}
\end{minipage}
\end{tabular}

\caption{(a) Quad well potential with four metastable sets and one-dimensional transition pathway. (b) Largest eigenvalues of the transfer operator $\mathcal{P}^t$. The spectral gap after the four dominant eigenvalues is clearly visible. (c) The four dominant eigenfunctions, encoding the hierarchy of the mixing processes between the four metastable sets.}
\label{fig:quadwell1}
\end{figure}

The first idea that comes to mind is to define the reaction coordiante directly as the dominant eigenfunctions (weighted by the stationary density for technical reasons):
\begin{equation} \label{eq:eigfunRC}
\xi(x) = \begin{pmatrix}
v_0(x) / \rho(x)\\
\vdots \\
v_d(x) / \rho(x)
\end{pmatrix}~.
\end{equation}
This reaction coordinate is indeed optimal, as was shown in Ref.~\onlinecite{Bittracher2017}. Indeed, the authors in Ref.~\cite{FGH14a} have also identified \eqref{eq:eigfunRC} as an ideal reaction coordinate, though only for a narrower sub-class of timescale-separated systems. However, there are two major practical disadvantages in choosing the eigenfunctions as reaction coordinates that ultimately prevent us from computing and using them:

\begin{enumerate}
\item The eigenproblem is global, and thus prohibitively expensive to solve numerically in high dimensions. If we wish to compute the value of an eigenfunction at only a single position in $\X$, we need an approximation of $\mathcal{P}^t$ that is accurate on all of $\X$. There have been attempts to mitigate this, but the conceptual problem remains.
\item The eigenfunction reaction coordinate often is redundant. In systems where the slow processes correspond to the transitions between $d$ metastable sets, i.e. $d$ potential wells, \eqref{eq:eigfunRC} would define a $d$-dimensional reaction coordinate. However, in practice, many of these potential wells often lie along the same transition path, and consequently the transitions between those wells would be describable by just a one-dimensional reaction coordinate. See the example in Figure \ref{fig:quadwell1} for an illustration.
\end{enumerate}

We will now reformulate criterion \eqref{eq:optcrit2} in a way that addresses the concerns above and at the same time makes it compatible with the transition path intuition of reaction coordinates.
Consider a reaction coordinate $\xi$ of some dimension $r\leq d$, fulfilling \eqref{eq:optcrit2}. Further, assume that for each starting point $x$, we can write the transition density $p^t(x,\cdot)$ as a linear combination of the eigenfunctions $v_i$:
$$
p^t(x,\cdot) = \sum_{i=0}^\infty d_i(x,t) v_i(\cdot),
$$
where $p^t(x,\cdot)$ denotes the $y$-dependent function $p^t(x,y)$ for all $y$.  
It can be shown (see Appendix \ref{ap:TransdensityDecomp}) that the prefactors are again connected to the eigenpairs: $d_i(x,t)=\lambda_i^t v_i(x)/\rho(x)$. As we still are interested only in long lag times $t$ where the non-dominant eigenvalues have already decayed, we can truncate the series:
$$
p^t(x,\cdot) \approx \sum_{i=0}^d \lambda_i^t \frac{v_i(x)}{\rho(x)}  v_i(\cdot)~.
$$
Finally, we use that $\xi$ fulfills the criterion \eqref{eq:optcrit2}, and that $\rho$ is the $0$-th dominant eigenfunction of $\mathcal{P}^t$ and get
\begin{equation}
\label{eq:transprobcrit}
p^t(x,\cdot) \approx \sum_{i=0}^d \lambda_i^t \frac{w_i\big(\xi(x)\big)}{w_0\big(\xi(x)\big)}  v_i(\cdot)~.
\end{equation}
The right hand side of this equation only depends on the reaction coordinate value $\xi(x)$, and not the full state space coordinate $x$. This means that the left hand side also can depend only on $\xi$.
Thus, in order for $\xi$ to be a good reaction coordinate, the transition density function $p^t(x,\cdot)$ must only depend on the $r$-dimensional $\xi(x)$, and not on the full $n$-dimensional $x$. We thus get the following equivalent criterion: $\xi$ is a good reaction coordinate if and only if
\begin{equation}
\label{eq:optcrit3}\tag{CII}
p^t(x,\cdot) \approx \tilde{p}^t\big(\xi(x),\cdot\big)
\end{equation}
for some function $\tilde{p}^t$, all $x$ and ``intermediate'' lag times~$t$. Intermediate here means that $t$ must be larger than the equilibration timescale of the fast processes, but can be chosen much smaller than the equilibration timescale of the slow processes. In terms of the implied timescales \eqref{eq:ImpliedTimescales} this writes $t_d > t > t_{d+1}$.

\subsection{Connection to Transition Path Theory}
\label{sec:ConnectionTPT}

Unlike \eqref{eq:optcrit2}, criterion \eqref{eq:optcrit3} now allows an interpretation in the context of Transition Path Theory (TPT). To be precise, we argue that the committor function, which is seen in TPT as the optimal reaction coordinate \cite{ElEtAl17}, fulfills criterion \eqref{eq:optcrit3}.

In a system with two metastable sets A and B, the forward committor function $q_A(x)$ is defined as the probability that the process $\bm{X}_t$ first visits $A$ rather than $B$ given the starting point $X_0=x$.
For a starting point outside the metastable sets, and for ``intermediate'' lag times t as required by \eqref{eq:optcrit3}, the probability to find the system in one of the metastable sets after time t is essentially~1, as the process quickly leaves the transition region. Moreover, the system equilibrates quickly inside the metastable sets.  Thus, the transition density essentially depends only on whether it is more likely to find the evolved system in $A$ or in $B$:
\begin{align*}
p^t(x,\cdot) &\approx c^t_A(x) \mathbbm{1}_A(\cdot)\rho(\cdot) + c^t_B(x)  \mathbbm{1}_B(\cdot)\rho(\cdot)~.
\intertext{Here, $\mathbbm{1}_A$ denotes the indicator function over $A$ and $c^t_A(x),~c^t_B(x)$ are the probabilities to find the evolved system in $A$ and $B$, respectively:}
c^t_A(x) &= \mathbf{Pr}\big[\bm{X}_t \in A~|~X_0=x\big]~,\\
c^t_B(x) &= \mathbf{Pr}\big[\bm{X}_t \in B~|~X_0=x\big] \approx 1-c^t_A(x)~.
\end{align*}
As we have chosen $t$ as \emph{intermediate}, i.e. so short that it is unlikely to leave a metastable set within time $t$ once it has been reached, $c^t_A(x)$ is essentially equal to the committor function. 
Thus we have
$$
p^t(x,\cdot) \approx q_A(x) \mathbbm{1}_A(\cdot)\rho(\cdot) + \big((1-q_A(x)\big)  \mathbbm{1}_B(\cdot)\rho(\cdot)~,
$$
where we see that the right hand side now only depends on $q_A(x)$, and not on the full value $x$.
With the function
$$
\tilde{p}^t(\xi,\cdot) := \xi \mathbbm{1}_A(\cdot) \rho(\cdot) + (1-\xi) \mathbbm{1}_B(\cdot) \rho(\cdot)~,
$$
 the reaction coordinate $\xi(x):=q_A(x)$ thus fulfills criterion $\eqref{eq:optcrit3}$. Our new criterion thus confirms the committor function as a good reaction coordinate in the sense of preserving the slow transition process between the metastable sets.

Note however that while the definition of committor functions depends on the existence (and the knowledge) of metastable states, Criterion \eqref{eq:optcrit3} can also be applied in systems where the slowest processes do not correspond to transitions between metastable states (such as systems with explicit timescale separation).
Criterion \eqref{eq:optcrit2} does not even require a spectral gap at all, i.e. reaction coordinates fulfilling \eqref{eq:optcrit2} will preserve the $d$ slowest processes even if the subsequent processes live on similar timescales.
Thus, our theory offers a much more general characterization of good reaction coordinates that however agrees with the concept of committor functions in the special cases where the latter is applicable.
What is more, the usage of committor functions as reaction coordinates is susceptible to the same computational problems as transfer operator eigenfunctions that were detailed in Section \ref{sec:criterion}.

The following example demonstrates that Criterion \eqref{eq:optcrit3} can formally distinguish ``intuitively good'' from ``intuitively bad'' reaction coordinates:
\begin{example} \label{ex:BananaPot}
 In Figure \ref{fig:quadwell2} we consider a diffusion process \eqref{eq:SmolDyn} in the curved double well potential
$$
V(x_1,x_2) = (x_1^2-1)^2 + 2(x_1^2+x_2-1)^2
$$ 
that was first analyzed in Ref.~\onlinecite{LeLe10} in the context of reaction coordinates. The inverse temperature was chosen as $\beta=0.5$. First, consider the one-dimensional reaction coordinate 
$$
\xi(x_1,x_2) = x_1 \exp(-2x_2)~.
$$
The transition pathway (here taken as the minimum energy pathway) is parametrized by $\xi$, i.e. no two points on the transition pathway take the same value under $\xi$. Further, the isolines (sets of constant value) of $\xi$ intersect the transiton pathway perpendicularly. $\xi$ was identified in Ref.~\onlinecite{LeLe10} as the ideal reaction coordinate, and can also be considered ``intuitively good'' from the standpoint of transition pathways. Note however that $\xi$ is not equal to the committor function.

On the other hand, the reaction coordinate
$$
\zeta(x_1,x_2) = x_2
$$
is obviously bad, as it does not parametrizes the transition pathway.

We can distinguish $\xi$ and $\zeta$, without any knowledge of the transition pathway, the metastable sets or the potential, by considering only the transition density functions $p^t(x,\cdot)$ along their isolines. We observe that for different starting points along any isoline of $\xi$, the densities $p^t(x,\cdot)$ for intermediate lag time $t$ look very similar.
That means that $p^t(x,\cdot)$ effectively only depends on $\xi(x)$. The same property does not hold for the bad reaction coordinate $\zeta$: here the densities $p^t(x,\cdot)$ differ substantially for starting points along a single isoline. In conclusion, $\xi$ fulfills Criterion \eqref{eq:optcrit3}, whereas $\zeta$ does not, i.e. criterion \eqref{eq:optcrit3} can distinguish good from bad reaction coordinates (in this example).
\begin{figure}
\begin{tabular}{m{.01\textwidth} m{.24\textwidth} m{.25\textwidth}}
&
\begin{center}(a)\end{center}
&
\begin{center}(b)\end{center}
\\
$\xi$
&
\includegraphics[scale=1]{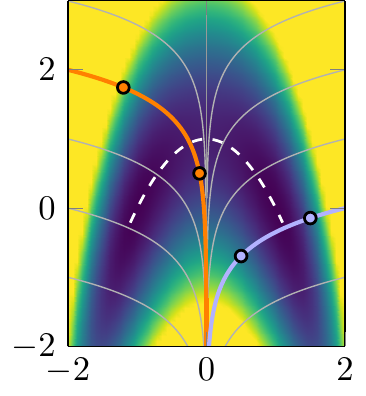}
&
\begin{minipage}{.24\textwidth}
\begin{minipage}{.48\textwidth}
\includegraphics[scale=1]{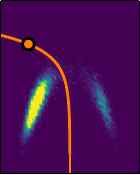}
\end{minipage}
\begin{minipage}{.48\textwidth}
\includegraphics[scale=1]{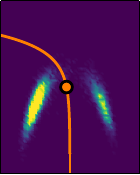}
\end{minipage}

\vspace{0.8em}
\begin{minipage}{.48\textwidth}
\includegraphics[scale=1]{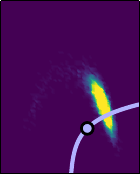}
\end{minipage}
\begin{minipage}{.48\textwidth}
\includegraphics[scale=1]{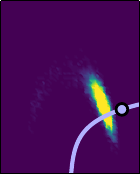}
\end{minipage}
\end{minipage}

\\
\hline
\\

$\zeta$
&
\includegraphics[scale=1]{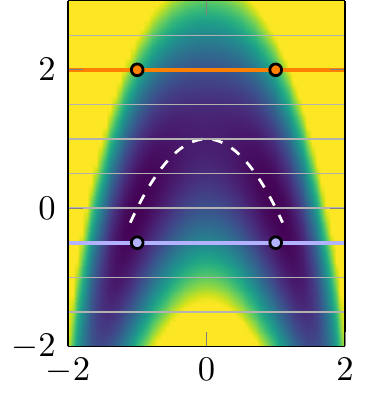}
&
\begin{minipage}{.24\textwidth}
\begin{minipage}{.48\textwidth}
\includegraphics[scale=1]{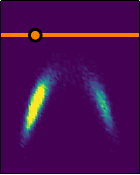}
\end{minipage}
\begin{minipage}{.48\textwidth}
\includegraphics[scale=1]{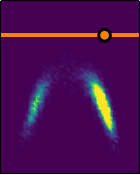}
\end{minipage}

\vspace{0.8em}
\begin{minipage}{.48\textwidth}
\includegraphics[scale=1]{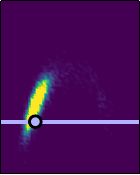}
\end{minipage}
\begin{minipage}{.48\textwidth}
\includegraphics[scale=1]{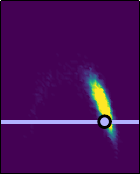}
\end{minipage}
\end{minipage}

\end{tabular}

\caption{(a) The curved double well potential, with the isolines of the reaction coordinates $\xi$ and $\zeta$ (grey lines) and the transition pathway (white dashed line). (b) transition density functions $p^t(x,\cdot)$ with different starting points $x$ on the highlighted isolines ($t=4$).}
\label{fig:quadwell2}
\end{figure}

\end{example}

\begin{summary*}
The equivalent criteria \eqref{eq:optcrit2} and \eqref{eq:optcrit3} allow for a rigorous characterization of good reaction coordinates such that the long time scales of the full molecular system are inherited by the projection of the dynamics onto the low-dimensional state space spanned by the reaction coordinates. At the same time, these criteria agree with but extend and refine the comprehension of good reaction coordinates that is pervasive in transition path theory. 
\end{summary*}

\section{Numerical computation of good reaction coordinates}
\label{sec:methods}

\subsection{The transition manifold}

From now on, we always assume $t$ to be an ``intermediate'' lag time as required by criterion \eqref{eq:optcrit3}.
This criterion implies that two transition density functions $p^t(x_1,\cdot)$ and $p^t(x_2,\cdot)$ are close to each other for two points $x_1,x_2$ of similar reaction coordinate value, even if $x_1$ and $x_2$ themselves are not close. We will now render this ``neighborhood relation'' of densities more precise and exploit it in order to efficiently compute good reaction coordinates.

For each state space point $x$, the transition density $p^t(x,\cdot)$ is a function in the infinite-dimensional function space $L^1$, i.e. the space of absolutely integrable functions. However, the insight that $p^t(x,\cdot)$ effectively depends only on $\xi(x)$, i.e. an $r$-dimensional coordinate, implies that the set of all transition density functions,
$$
\mathbb{M} = \{ p^t(x,\cdot)~|~x\in \X \}~,
$$
effectively forms an only $r$-dimensional manifold in this function space. In the common case of $r=1$, $\mathbb{M}$ is effectively a curve in $L^1$. We call $\mathbb{M}$ the \emph{transition manifold} of the system. 

\begin{remark*}
While there is a connection between transition path theory and transition manifolds as shown in Section \ref{sec:criterion}, to the best of the authors' knowledge, there is no formal equivalence between the transition manifold and any existing definition of transition pathway.
\end{remark*}

Assume now that we are able find any parametrization of $\mathbb{M}$, i.e. a smooth invertible function $\mathcal{E}: \mathbb{M}\rightarrow \mathbb{R}^r$. Then one can show\cite{Bittracher2017} that the reaction coordinate defined as
\begin{equation}
\label{eq:RC}
\xi(x) := \mathcal{E}\big( p^t(x,\cdot)\big)
\end{equation}
fulfills the criterion \eqref{eq:optcrit3} (or equivalently \eqref{eq:optcrit2}). This is the reaction coordinate we will ultimately compute numerically.

\subsection{Embedding of the transition manifold}
\label{sec:TMembedding}

In order to find a parametrization of the transition manifold $\mathbb{M}$, we employ the general-purpose \emph{Diffusion Maps} manifold learning algorithm\cite{CoLa06,singer2008non}. Explaining the algorithm in detail would go well beyond the scope of this article, so we only coarsely sketch its usage: Let a sufficiently large collection of data points $\{z_1,\ldots,z_M\}$ on or near a manifold in some vector space be given. The algorithm then detects the dimension $r$ of this manifold, and returns for each data point $z$ an $r$-dimensional vector $(\mathcal{E}_1(z),\ldots,\mathcal{E}_r(z))^\intercal$ that represents a parametization $\mathcal{E}$ of the manifold, evaluated at $z$. 
Application of Diffusion Maps requires the choice of a certain kernel bandwidth parameter that essentially determines what distance should be considered ``far away''. We assume from now on that this parameter can be chosen reliably, an optimal strategy has been detailed in Ref.~\onlinecite{Sin06}.

The Diffusion Maps algorithm in principle works in arbitrary metric spaces, as it only requires an appropriate notion of distance between data points. We will however not attempt to parametrize the transition manifold directly in $L^1$, as the calculation of distances between $L^1$ functions is numerically costly. Instead, we will first \emph{embed} the transition manifold $\mathbb{M}$ into a Euclidean space, and use the standard Euclidean distance there.

\begin{figure*}
\includegraphics[scale=1]{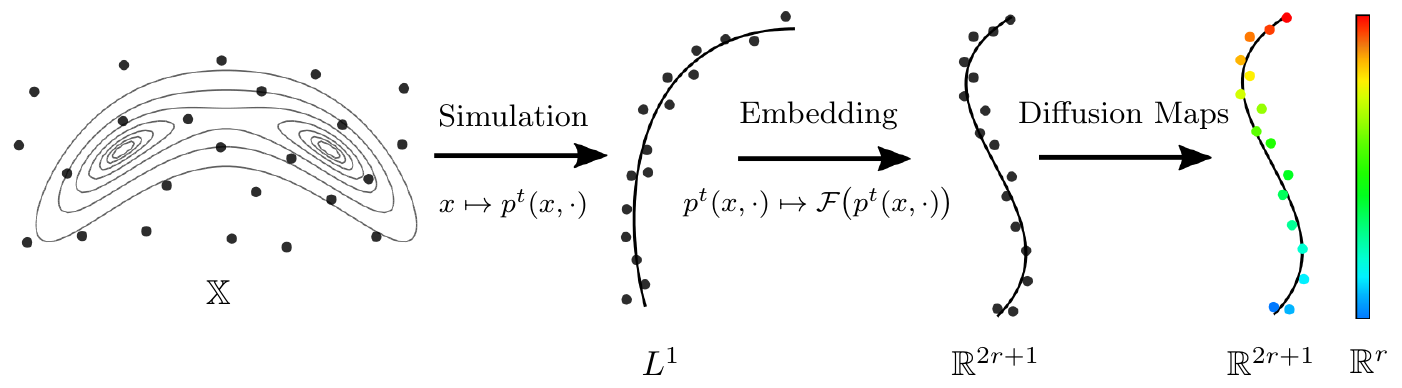}
\caption{Illustration of the complete transition manifold embedding procedure as implicitly performed in Algorithm~\ref{algo:PointwiseRC}.}
\label{fig:EmbeddingSketch}
\end{figure*}

Surprisingly, constructing such an embedding requires virtually no knowledge about $\mathbb{M}$. Let $\mathcal{F}:L^1 \rightarrow \mathbb{R}^{q}$ be an \emph{arbitrarily-chosen} map from the function space $L^1$ to the Euclidean space of dimension $2r+1$ (or greater), where $r$ is the dimension of the transition manifold. Then---slightly simpliefied---the famous Whitney embedding theorem\cite{Whi36,HuKa99} states that for any such $\mathcal{F}$ the probability for $\mathcal{F}(\mathbb{M})$ again being an $r$-dimensional manifold in $\mathbb{R}^{2r+1}$ is \emph{exactly one}. For the purpose of this article, this means that we can effectively choose $\mathcal{F}$ randomly---if only its image dimension is large enough---and be sure that the manifold structure of $\mathbb{M}$ gets preserved under~$\mathcal{F}$. We can then compute a parametrization of the \emph{embedded} manifold $\mathcal{F}(\mathbb{M})$ using the Diffusion Maps algorithm, which then corresponds to a parametrization of the original manifold $\mathbb{M}$. A sketch of the overall embedding procedure is shown in Figure \ref{fig:EmbeddingSketch}.

Specifically, we will work with the $2r+1$ embedding functions
\begin{equation}
\label{eq:embedding}
\mathcal{F}_i\big(p^t(x,\cdot)\big) = \int_\mathbb{X} \eta_i(y) p^t(x,y)~dy~,
\end{equation}
with linear observables
\begin{equation}
\label{eq:observables}
\eta_i(x) =  a_{i1} x_1 + \ldots + a_{in} x_n
\end{equation}
where the factors $a_{ij}$ are chosen randomly (e.g. uniformly drawn from the interval $[0,1]$). Note however that we have great freedom in the choice of the functions $\eta_i$. Linear functions were chosen simply out of convenience.

We see immediately that by this choice of the embedding, the embedded density \eqref{eq:embedding} then is the Koopman operator \eqref{eq:koopmanoperator} applied to $\eta$,
i.e. the expectation value of $\eta$ under the evolved dynamics:
\begin{equation}
\label{eq:embedding2}
\mathcal{F}\big(p^t(x,\cdot)\big) = \mathbb{E}\big[ \eta(\bm{X}_t)~|~X_0=x\big]~.
\end{equation}
The right hand side can now be computed numerically by a simple Monte Carlo sampling procedure. Let $\Phi^t_{j}(x),~j=1,\ldots,M$ denote the endpoints of $M$ independent trajectories of length $t$, all starting in $x$.
Then 
$$
\mathcal{F}\big(p^t(x,\cdot)\big) \approx \frac{1}{M} \sum_{j=1}^M \eta\big(\Phi^t_j(x)\big)~.
$$

Thus, we create the data points in $\mathbb{R}^{2r+1}$ that we apply the Diffusion maps algorithm to as follows:
\begin{algorithm}[H]
\floatname{algorithm}{Algorithm}
\caption{Point-wise computation of the reaction coordinate}
\label{algo:PointwiseRC}
\begin{algorithmic}[1]
\STATE Choose $M$ points $\{x_1,\ldots,x_L\}$ that cover the relevant parts of state space, i.e. the metastable sets and the transition regions.
\STATE Choose the factors $a_{ij}$ in \eqref{eq:observables}, e.g. uniformly randomly in~$[0,1]$.
\STATE For each $x_i$, simulate $M$ independent trajectories of length $t$. Let the end points be denoted by $\Phi^t_j(x_i)$.
\STATE Compute the data points in $\mathbb{R}^{2r+1}$ as
$$
z_i \gets \frac{1}{M} \sum_{j=1}^M \eta\big(\Phi^t_j(x_i)\big)~.
$$
\STATE Apply the Diffusion Maps algorithm to $\{z_1,\ldots,z_L\}$.
\ENSURE Approximation to the $r$-dimensional reaction coordinate \eqref{eq:RC}, evaluated at the points $\{x_1,\ldots,x_L\}$, i.e. 
$$
\{\xi(x_1),\ldots,\xi(x_L)\}~.
$$
\end{algorithmic}
\end{algorithm}

\begin{remark*}
The above algorithm requires the knowledge of two intrinsic parameters of the system: 1) the ``intermediate'' lag time $t$, in order to simulate trajectories of the right length, and 2) the expected dimension $r$ of the reaction coordinate, in order to choose the right number of embedding observables.
For both quantities, rough estimates can be used in practice.

The weak requirement $t_\text{slow} > t > t_\text{fast}$ on the lag time $t$ permits a high tolerance with respect to numerical errors. Thus, rough Markov models can for example be used to estimate the timescales. Also, in real-live chemical systems, one often has a general idea about the nature of the fast and slow processes (e.g. whether one is interested in the re-configuration of individual dihedral angles or the forming of higher-level structures) that can guide the choice.

For the dimension $r$, an iterative procedure can be used: First start with a low estimate for $r$ (e.g., $r=1$) and perform Algorithm \ref{algo:PointwiseRC}. 
If the chosen $r$ was equal to or higher than the correct dimension of the transition manifold, the Diffusion Maps algorithm should detect an $r$- or lower-dimensional manifold in the embedded data points. If it fails to do so, increase $r$ by choosing additional observables and restart the embedding procedure. This strategy generates only little overhead, as the simulations $\Phi_j^t(x_i)$ and the previously embedded points can be reused.

Alternatively, assuming the rough Markov model mentioned above can correctly identify the number $d$ of dominant timescales, this can be used as an upper bound for~ $r$. Even if $d$ vastly overestimates $r$, the final reaction coordinates $\xi$ (after application of the Diffusion Maps algorithm) will have the correct low dimension $r$.
\end{remark*}

\section{Galerkin approach for computing reaction coordinates}
\label{sec:Galerkin}

As we have seen, the transition manifold-based reaction coordinate \eqref{eq:RC} fulfills rigorous optimality criteria regarding the preservation of the long timescales and being of the smallest possible dimension.
Unfortunately, the above algorithm to compute it has two major practical shortcomings:
\begin{enumerate}
\item $\xi$ can only be computed \emph{pointwise} and has no closed analytic form. For every new evaluation point many numerical MD simulations have to be started. Further, the evaluation points have to be chosen in regions relevant to the slow transition processes (i.e., in the transition regions and metastable sets), which is a non-trivial task, especially in high-dimensional systems.
\item The computation of $\xi$ is based on multiple short, instead of one long MD simulation. Although this can also be seen as an advantage, the way modern MD software works often favors the simulation of single long trajectories. Further, there is a vast archive of already pre-computed trajectories for many interesting metastable molecular systems. If this data could be used to compute $\xi$, those systems could be coarse-grained with minimal effort.
\end{enumerate}

In the following we will thus describe a Galerkin discretization of the embedding function \eqref{eq:embedding}. 
Importantly, this discretization will be very similar to the discretization of the dominant transfer operator eigenfunctions performed in MSM analysis, further emphasizing the close connection of the methods.
Moreover, this will allow us to calculate our reaction coordinates from the same data sources also used in MSM building, and utilize a wide range of analogous discretization techniques.

\subsection{Galerkin approximation of reaction coordinates}

We first write the embedded density \eqref{eq:embedding2} directly as a function of the starting point $x$:
$$
\tilde{\xi}(x) = \mathbb{E}\big[ \eta(\bm{X}_t)~|~X_0=x\big]~.
$$
We make the weak assumption that all the components of $\tilde{\xi}$ are square-integrable with respect to the stationary density, i.e. $\tilde{\xi}$ lies in the function space $L^2_\rho$ with inner product
\begin{equation}
\label{eq:rhoinnerproduct}
\langle f,g\rangle_\rho = \int_\X f(x) g(x) \rho(x)~dx~.
\end{equation}

\begin{remark*}
The  function $\tilde{\xi}$ can already be understood as a $2r+1$-dimensional reaction coordinate; i.e. we could in theory accept a reaction coordinate with higher than optimal dimension in order to save us the application of the Diffusion Maps algorithm. Thus, we will refer to $\tilde{\xi}$ as the ``pre-reaction coordinate''.
\end{remark*}

We now discretize $\tilde{\xi}$ using a Galerkin approximation\cite{SchSa13}, i.e.  we seek the function $\tilde{\xi}_N$ inside a finite-dimensional function space $\mathcal{V}_N$ that best approximates~$\tilde{\xi}$.
Classical choices of the ansatz space $\mathcal{V}_N$ are, for example, the space of all polynomials over $\X$ up to a certain degree, the space consisting of $N$ characteristic functions over a finite partition of $\X$, or some other finite element space.
The Galerkin approximation is performed independently on the $2r+1$ individual components of $\tilde{\xi}$. However, we will omit the subscripts in order to help readability and simply treat $\tilde{\xi}$ and $\eta$ as one-dimensional functions for the remainder of this section.

Let $\{\varphi_1,\ldots,\varphi_N\}$ be a basis of $\mathcal{V}_N$. Then the Galerkin approximation $\tilde{\xi}_N$ has the following closed form:
\begin{equation}\label{eq:Galerkin_xi}
\tilde{\xi}_N(x) \approx \sum_{k,j=1}^N \varphi_j(x) (S^{-1})_{kj} \sum_{l=1}^NT_{kl}c_l~,
\end{equation}
with the \emph{Gram matrix}
\begin{align*}
S_{kj} &= \langle \varphi_k,\varphi_j\rangle_\rho~,\\
\shortintertext{the \emph{transition matrix}}
T_{kl} &= \langle \mathcal{P}^t \varphi_k,\varphi_l\rangle_\rho
\intertext{where $\mathcal{P}^t$ is again the transfer operator of the system, and the factors}
c_l &= \sum_{k=1}^N (S^{-1})_{kl} \langle\eta,\varphi_k\rangle_\rho~,
\end{align*}
where $\eta$ is the randomly-chosen observable from \eqref{eq:embedding}. The precise derivation of equation \eqref{eq:Galerkin_xi} is given in Appendix \ref{ap:GalerkinRCFormula}. The quantities $S$, $T$ and $c$ can now be computed numerically,
and thus $\tilde{\xi}_N$ can be evaluated at any state space point $x$.

\begin{remark*}
The exact the matrices $S$ and $T$ are also commonly found at the heart of methods that aim to reconstruct long-term dynamics directly via the transfer operator eigenfunctions\cite{Schuette1999, tika2013}, such as Markov State Models. The Galerkin approximation of $\tilde{\xi}$ is thus applicable whenever those methods are.
\end{remark*}

%
%

\subsection{Data-based computation of the transition matrix.}

The entries of the transition matrix $T$ and Gram matrix $S$ can now be approximated based on simulation data.
Consider two sets of data points on $\X$,
\begin{equation}
\label{eq:DataSets}
\mathbb{X}_M = \{x_1,\ldots,x_M\}~,\quad \mathbb{Y}_M = \{y_1,\ldots,y_M\}~,
\end{equation}
where $\mathbb{X}_M$ samples the stationary density $\rho$, \todo{Andreas: is it enough if $\mathbb{X}_M$ only samples the quasistationary measure $\hat{\mu}$ of $\hat{\mathbb{X}}$ ?)} and  $\mathbb{Y}_M\subset\mathbb{X}$ is the time-$t$ evolution of $\mathbb{X}_M$ under the dynamics. To be precise, $y_i=\Phi^tx_i$, with $t$ being again the ``intermediate'' lag time. This data can for example be obtained from a single equilibrated numerical trajectory of step size $\tau$ (assuming that $t$ is a multiple of $\tau$),
\begin{equation}
\begin{aligned}
\label{eq:TrajDataSets}
\mathbb{X}_M &= \{x_0, \Phi^\tau x_0,\ldots,\Phi^{(M-1)\tau}x_0\}~,\\ \mathbb{Y}_M &= \{\Phi^tx_0, \Phi^{\tau+t}x_0,\ldots,\Phi^{(M-1)\tau +t}x_0\}~,
\end{aligned}
\end{equation}
or the concatenation of multiple trajectories that \emph{together} sufficiently sample $\rho$.
Alternatively, $\mathbb{X}_M$ could be the output of an enhanced sampling algorithm, such as Markov chain Monte Carlo methods \cite{gilks1995markov}, and $\mathbb{Y}_M$ the endpoints of individual trajectories starting in $\mathbb{X}_M$.

As frequently used in the Markov state approach\cite{KKS16}, the inner product $\langle \cdot,\cdot\rangle_\rho$ can be approximated from $\rho$-distributed data via Monte Carlo quadrature. $T$ and $S$ can thus be approximated as
\begin{align*}
T_{kl} &\approx \frac{1}{M} \sum_{j=1}^M\varphi_k(x_j)\varphi_l(y_j)~, \\
S_{kl} &\approx \frac{1}{M} \sum_{j=1}^M\varphi_k(x_j)\varphi_l(x_j)~.
\end{align*}
Moreover, the factors $c_l$ become
$$
c_l \approx \sum_{k=1}^N (S^{-1})_{kl} \Big( \frac{1}{M}\sum_{j=1}^M \eta(x_j) \varphi_k(x_j) \Big)~.
$$

Subsequently, \eqref{eq:Galerkin_xi} can be evaluated at arbitrary state space points without significant additional costs. Choosing evaluation points $x_i$ that again cover the relevant parts of state space, for example a subsample of the data points $\X_M$, we can apply the Diffusion maps algorithm to the embedded points $\{\tilde{\xi}_N(x_1),\ldots,\tilde{\xi}_N(x_L)\}$ and again extract the final $r$-dimensional reaction coordinate. Algorithm \ref{algo:GalerkinRC} shows an accordingly  modified version of Algorithm \ref{algo:PointwiseRC}.
\begin{algorithm}[H]
\floatname{algorithm}{Algorithm}
\caption{Galerkin-based computation of the reaction coordinate}
\label{algo:GalerkinRC}
\begin{algorithmic}[1]
\REQUIRE{Data sets $\X_M,\mathbb{Y}_M$ as in \eqref{eq:DataSets}.}
\STATE Choose a Galerkin basis $\{\varphi_1,\ldots,\varphi_N\}$ that adequately approximates smooth functions over the relevant parts of state space.
\STATE Choose the factors $a_{ij}$ in \eqref{eq:observables}, e.g. uniformly randomly in~$[0,1]$.
\STATE Compute the matrices $T,S$ and the vector $c$ via
\begin{align*}
T_{kl} &\gets \frac{1}{M} \sum_{j=1}^M\varphi_k(x_j)\varphi_l(y_j)~, \\
S_{kl} &\gets \frac{1}{M} \sum_{j=1}^M\varphi_k(x_j)\varphi_l(x_j)~, \\
c_l &\gets \sum_{k=1}^N (S^{-1})_{kl} \Big( \frac{1}{M}\sum_{j=1}^M \eta(x_j) \varphi_k(x_j) \Big)~.
\end{align*} \STATE Choose $L$ evaluation points $\{x_1,\ldots,x_L\}$ that cover the relevant parts of state space, i.e. the metastable sets and the transition regions.
\STATE Compute the data points in $\mathbb{R}^{2r+1}$ as
\begin{equation*}
z_i \gets \sum_{k,j=1}^N \varphi_j(x_i) (S^{-1})_{kj} \sum_{l=1}^NT_{kl}c_l~,
\end{equation*}
\STATE Apply the Diffusion Maps algorithm to $\{z_i,\ldots,z_M\}$.
\ENSURE Approximation to the $r$-dimensional reaction coordinate \eqref{eq:RC}, evaluated at the points $\{x_1,\ldots,x_L\}$, i.e. 
$$\
\{\xi(x_1),\ldots,\xi(x_L)\}~.
$$
\end{algorithmic}
\end{algorithm}

\begin{remark*}
Another advantage of Algorithm \ref{algo:GalerkinRC} is that, when adding a new evaluation point $x_{L+1}$, no new simulations have to be started. Only the Diffusion Map algorithm has to be re-applied to the now extended embedded points $\{z_1,\ldots,z_{L+1}\}$.
\end{remark*}

\subsection{Implementation: Voronoi-based Galerkin approximation}
\label{sec:VoronoiGalerkin}

For Markov State Model construction, there exists an extensive collection of elaborate Galerkin basis sets that have been successfully applied to real-world biomolecular systems, and all of them can in principle be used to approximate the reaction coordinate $\xi$.
Examples are hierarchical wavelet bases \cite{JuKo09}, meshfree basis functions based on Shepards approach \cite{FaBuWe13,WeberFackeldeySchuette2017}, and specialized problem-adapted basis sets, such as a tensor basis for peptide chains \cite{ViNoKe15}.
In this section, we detail a simple, yet practical algorithm that constructs a particular meshfree ansatz space directly from the available simulation data. Similar basis functions have been explored in the context of MSMs in Ref.~\onlinecite{Web06}.

Let~$\{A_1,\ldots,A_N\}$ be sets that partition $\mathbb{X}$, i.e. $
\bigcup_i A_i = \mathbb{X}$ and $A_i\cap A_j = \varnothing,~i\neq j$.
 Choosing the indicator functions over the sets $A_i$,
$$
\varphi_k(x) := \begin{cases}
1 & x\in A_k\\
0 & \text{otherwise}
\end{cases}~,
$$ 
as the basis of $\mathcal{V}_N$, the entry $T_{kl}$ of the transition matrix is effectively just the relative number of transitions from set $A_k$ to set $A_l$ within the data sets~$\mathbb{X}_M,~\mathbb{Y}_M$. The Gram matrix is diagonal, with $S_{kk}$ being the relative number of data points in $\X_M$ that lie in $A_k$. This partition-based Galerkin approximation of the transfer operator is known as Ulam's method in the MSM literature\cite{Ulam1960}.

The evaluation of $\tilde{\xi}_N$ at a specific point $x\in A_k$, then becomes
$$
\tilde{\xi}_N(x) =  S_{kk}^{-1}\sum_{l=1}^N T_{kl} c_l~.
$$

\subsubsection*{Choice of the partition sets}
Choosing the partition sets naively, for example as a regular box grid, invokes the infamous \emph{curse of dimensionality}, as the number of boxes rises exponentially with the system's dimension. We thus propose a partition into grid-free Voronoi cells $\{A_1, \ldots, A_N\}$ with center points adapted to the dynamical data $\X_M$. With this, we will also be able to avoid the explicit construction (and storage) of the transition matrix.

Our objective is to approximate $\tilde{\xi}$ in the region of state space that is covered with the available data points $\mathbb{X}_M$. The question is then how the Voronoi centers $E=\{e_1,\ldots,e_N\}\subset \mathbb{X}$ should be chosen in order to achieve this. In the following, we demonstrate that two different criteria on the approximation quality of $\tilde \xi$ lead to two different algorithms for selecting the Voronoi centers.

\paragraph{Minimizing the $L^2$ error.} Since $\tilde\xi\in L^2_\rho$, we may ask to minimize the error
\begin{equation}
\label{eq:L2Error}
\|\tilde{\xi} - \tilde{\xi}_N\|_\rho \overset{!}{=} \min_{\{e_1,\ldots,e_N\}\subset \mathbb{X}}~,
\end{equation}
where $\|\cdot\|_\rho$ is the norm induced by the inner product \eqref{eq:rhoinnerproduct}.
In Appendix \ref{ap:VoronoiCenters} we show that under weak assumptions, this error is minimized by choosing as the Voronoi centers the output of the $k$-means clustering algorithm \cite{Lloyd82} applied to the data $\X_M$ with $k=N$. $k$-means is highly scalable for both large amounts of clusters $N$ and a large number of data points $M$, and is readily available in many software packages.

\paragraph{Minimizing uniform error.} Thinking of $\tilde \xi$ as an observable, it is natural to minimize the \emph{uniform} observable error
\begin{equation}
\label{eq:UniformError}
\|\tilde{\xi} - \tilde{\xi}_N\|_\infty \overset{!}{=} \min_{\{e_1,\ldots,e_N\}\subset \mathbb{X}}~.
\end{equation}
In Appendix \ref{ap:VoronoiCenters} we show that, again under weak assumptions,
 the minimum is achieved if the centers cover the region of $\X$ where data is available evenly such that the Voronoi cells all have similar diameters. 
This problem is related to \emph{Poisson disk-} or \emph{blue noise (sub)sampling} in computer vision\cite{Bridson2007}.
The following \emph{picking algorithm}\cite{WeberFackeldeySchuette2017} computes an approximately equidistant subsample of~$\X_M$:

\begin{algorithm}[H]
\floatname{algorithm}{Algorithm}
\caption{Picking algorithm}
\label{algo:PoissonSampling}
\begin{algorithmic}[1]
\REQUIRE{$\mathbb{X}_M$, N}
\STATE $e_1 \gets$ random point from $\mathbb{X}_M$
\FOR {$j=2,\ldots,N$}
	\STATE \multiline{pick the point with the maximum distance from the previous points:}
	
	 $\displaystyle e_j \gets \argmax_{x\in\mathbb{X}_M} \min_{i=1,\ldots,j-1} \|x-e_i\|$
\ENDFOR
\ENSURE{Voronoi centers $E=\{e_1,\ldots,e_N\}$}
\end{algorithmic}
\end{algorithm}

In conclusion, minimizing the $L^2$ error of $\tilde \xi$ leads to $k$-means clustering as an algorithm for picking the Voronoi centers while minimizing the uniform error of $\tilde \xi$ leads to the farthest point picking algorithm \ref{algo:PoissonSampling}. In Section \ref{sec:examples} we compare both alternatives. In general $k$-means will lead to denser Voronoi cells in metastable regions, while algorithm \ref{algo:PoissonSampling} will lead to evenly sized Voronoi cells.

In order to compute $T,S$ and $c$, the data points from $\X_M$ and $\mathbb{Y}_M$ have to be assigned to their respective partition set. In the case of Voronoi cells this is easily done by a nearest point search between $\X_M$ and $E$, and $\mathbb{Y}_M$ and $E$, respectively.

\begin{summary*}
The criteria \eqref{eq:optcrit2} and \eqref{eq:optcrit3} and the concept of transition manifolds offer a new perspective on optimal reaction coordinates.
Reaction coordinates that fulfill these criteria can in fact be computed using the same data sources and state space discretization techniques as classical MSMs which means that the entire machinery invented for building MSMs can be utilized for their computation.
\end{summary*}

\section{Examples} \label{sec:examples}

\subsection{Curved double well potential}
\label{sec:BananaPot}

As our first demonstration, we compute the reaction coordinate of the simple curved double-well potential from Example \ref{ex:BananaPot} using Algorithm \ref{algo:GalerkinRC}. It will allow us to visualize the (embedded) transition manifold and compare the computed reaction coordinate with the minimum energy pathway and the committor function.

In this low-dimensional example, the relaxation timescales associated with the slowest processses of the full system can be computed numerically, see Table \ref{tab:BananaTimescales}. They were computed by a sufficiently fine approximation of the transfer operator, computation of its eigenvalues and using formula \eqref{eq:ImpliedTimescales}.

As expected, the system is timescale-separated, with the single slow timescale representing the mean expected waiting time\cite{bianchi2012} for a single transition between the two wells. The lag time $t=2$ falls in between the slow and fast timescales, so we use it as the ``intermediate'' lag time for Algorithm~\ref{algo:GalerkinRC}. 
Moreover, we assume the dimension $r=1$ of the transition manifold to be known.

As source of dynamical data, we utilize a single well-equilibrated trajectory 
$$
\{x_0,\Phi^\tau x_0,\Phi^{2\tau}x_0,\ldots\}
$$
of the dynamics with step size $\tau=10^{-2}$ and overall $2\cdot 10^7$ steps. This trajectory is used to construct the data sets $\X_M,~\mathbb{Y}_M$ via formula \eqref{eq:TrajDataSets}.

We partition the interesting region of $\mathbb{R}^2$ into 1000 Voronoi cells. The characteristic functions over the cells then form the Galerkin basis for Algorithm \ref{algo:GalerkinRC}, as detailed in Section \ref{sec:VoronoiGalerkin}. The centers of the cells can be chosen using either the k-means algorithm or the picking algorithm (Algorithm \ref{algo:PoissonSampling}); we compare both methods in the following. As the evaluation points $\{x_1,\ldots,x_L\}$ that are required for Algorithm \ref{algo:GalerkinRC}, we simply re-use the 1000 Voronoi center points, as they already cover the interesting state space regions.
 
\subsubsection*{Results}

\begin{figure*}
\centering
\includegraphics[scale=.7]{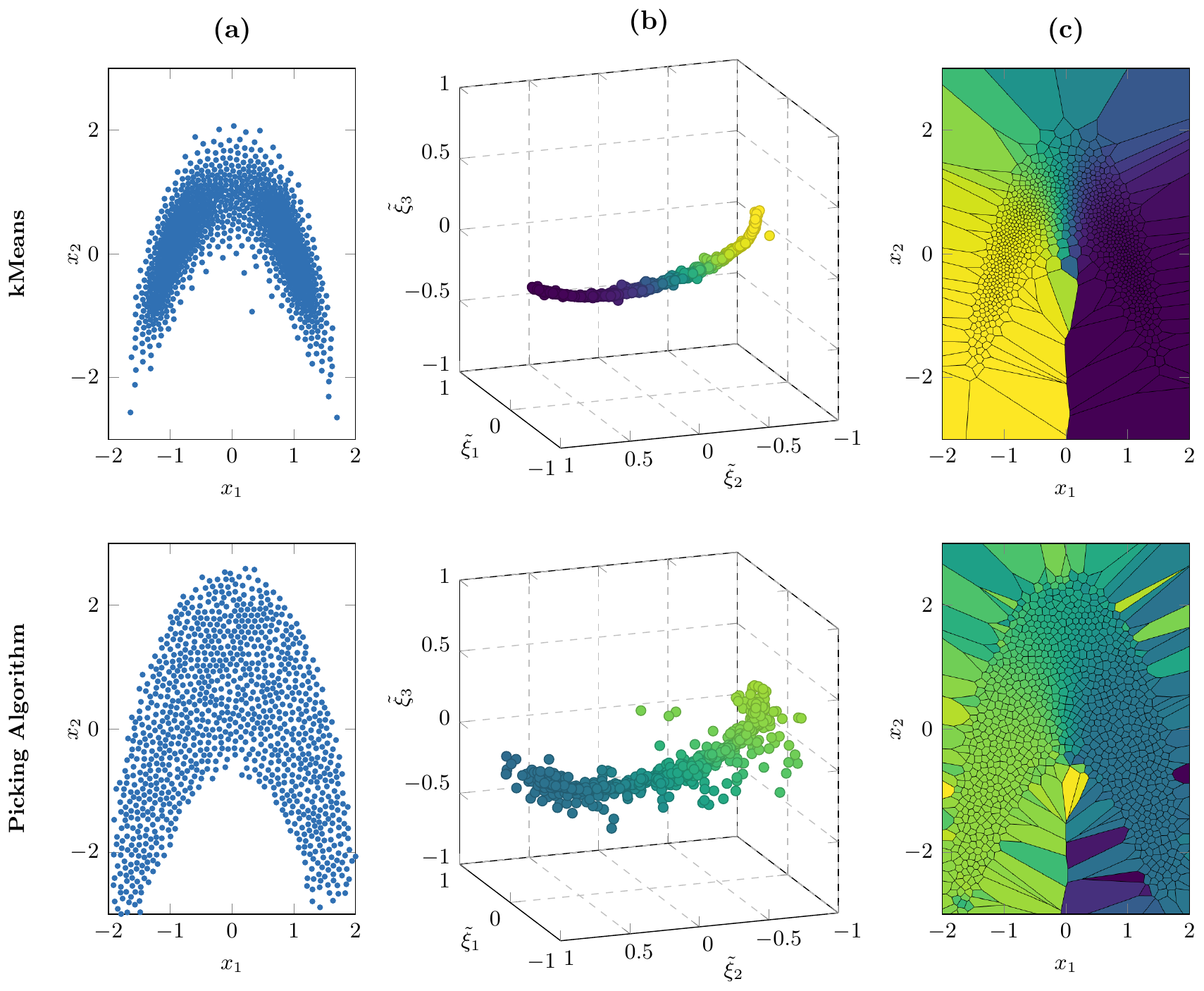}\hspace{2mm}
\raisebox{28mm}{\includegraphics[scale=.8]{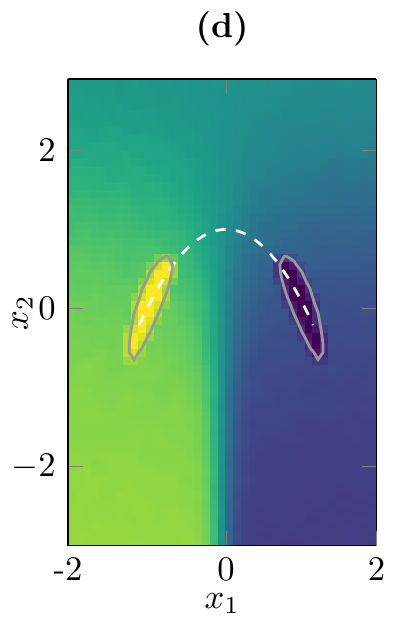}}
\caption{(a) Voronoi center points, computed with k-means clustering and the picking algorithm, respectively. (b) The three components of the computed reaction coordinate $\tilde{\xi}_N$, evaluated at the respective Voronoi centers. The output of the Diffusion Maps algorithm, which is used as the final one-dimensional reaction coordinate $\xi$, is used as color map. (c) $\xi$ as a function defined everywhere on $\mathbb{X}_M$. (d) Numerical approximation of the committor function between the two grey-framed metastable sets.}
\label{fig:BananaRC}
\end{figure*}

Figure \ref{fig:BananaRC} (a) shows the computed Voronoi center points. While the points based on the picking algorithm cover $\X_M$ evenly (by construction), the k-means-based center points appear to emphasize the metastable regions and slightly under-sample the transition regions.

Figure \ref{fig:BananaRC} (b) shows the approximation to $\tilde{\xi}$ evaluated at $\{x_1,\ldots,x_L\}$, computed via \eqref{eq:Galerkin_xi}. These are the $2r+1$-dimensional data points $z_i$ in Algorithm \ref{algo:GalerkinRC}. 
The points quite obviously concentrate around a one-dimensional manifold. This is the embedding of the transition manifold $\mathbb{M}$ into $\mathbb{R}^3$ via \eqref{eq:embedding}.

The Diffusion Maps algorithm applied to the points indeed finds the correct dimension $r=1$ of the embedded manifold and parametrizes it. The coloring in Figure \ref{fig:BananaRC} (b) indicates the value of the one-dimensional parametrization at the respective embedded point. This is also the value of the final reaction coordinate at the respective evaluation point, i.e. $\xi(x_i)$.
Assigning this value to the whole Voronoi cell that $x_i$ belongs to yields the final reaction coordinate $\xi$ that is defined in all $\mathbb{R}^2$, shown as the coloring in Figure \ref{fig:BananaRC} (c).

$\xi$ clearly parametrizes the minimum energy pathway, with a smooth gradient in the transition region. Moreover, $\xi$ qualitatively resembles the system's committor function that is shown in Figure \ref{fig:BananaRC} (d).

\subsubsection*{Timescale analysis}

To quantitatively verify the quality of the computed reaction coordinate, we compare the timescales of the full process $\bm{X}_t$ and the projected process $\xi(\bm{X}_t)$. Note that this is equivalent to comparing the dominant eigenvalues \eqref{eq:optcrit1}, i.e. a necessary condition for the criterion \eqref{eq:optcrit2}.
In fact, the timescales of the projected process were computed by first approximating the eigenvalues of the projected transfer operator (using the projected trajectory 
$\big\{\xi(x_0),\xi(\Phi^\tau x_0),\xi(\Phi^{2\tau} x_0),\ldots\big\}$) and then again using formula~\eqref{eq:ImpliedTimescales}.

Table \ref{tab:BananaTimescales} shows that our Galerkin-approximated reaction coordinate $\xi$ approximates the dominant timescale $t_1$ of the full system very well, both for Voronoi centers chosen by the k-means- and the picking algorithm. In fact, even the non-dominant timescales $t_2,t_3,\ldots$ are reproduced quite well, even though our theory only holds for the dominant timescales.
Compared to the naively-chosen reaction coordinate, $\zeta(x_1,x_2)=x_1$, our approximation error is noticeably lower, although $\zeta$ still preserves the timescales surprisingly well.

\begin{table}
dominant timescales
\vspace{.5em}

\setlength{\tabcolsep}{0.5em}
\begin{tabular}{r | c c c}	
          & $t_1$     & $t_2$     & $t_3$      \\
    \hline\addlinespace
    \textbf{\smash{full system}} 					& 5.9332 	 & 0.9021 	& 0.6031		\\
    \textbf{\smash{${\xi}$, k-means alg.}}     		& 5.8899     & 0.8615    & 0.5625     \\
    \textbf{\smash{${\xi}$, picking alg.}}     		& 5.9034     & 0.8789     & 0.5838     \\
    \textbf{\smash{$\zeta(x_1,x_2)=x_1$}}	    		& 5.7130     & 0.7964     & 0.5380     \\
    \bottomrule
\end{tabular}%

\caption{Dominant implied timescales of the full double well system, and the system projected onto different reaction coordinates. The zero-th timescale is $t_0=\infty$ in all four cases.}
\label{tab:BananaTimescales}

\end{table}

\subsection{Alanine dipeptide}

We demonstrate that with Algorithm \ref{algo:GalerkinRC}, one can successfully use longtime simulation data to identify quantitatively good reaction coordinates in realistic molecular systems, that the resulting reaction coordinates are interpretable chemically, and that the reaction coordinates can be used to quantitatively restore the information about the long-time transition processes (in form of the transfer operator eigenfunctions).


\begin{figure}
\centering

\setlength{\tabcolsep}{2pt}

\begin{tabular}{c}
\footnotesize \textbf{(a)} \\
\includegraphics[width=.29\textwidth]{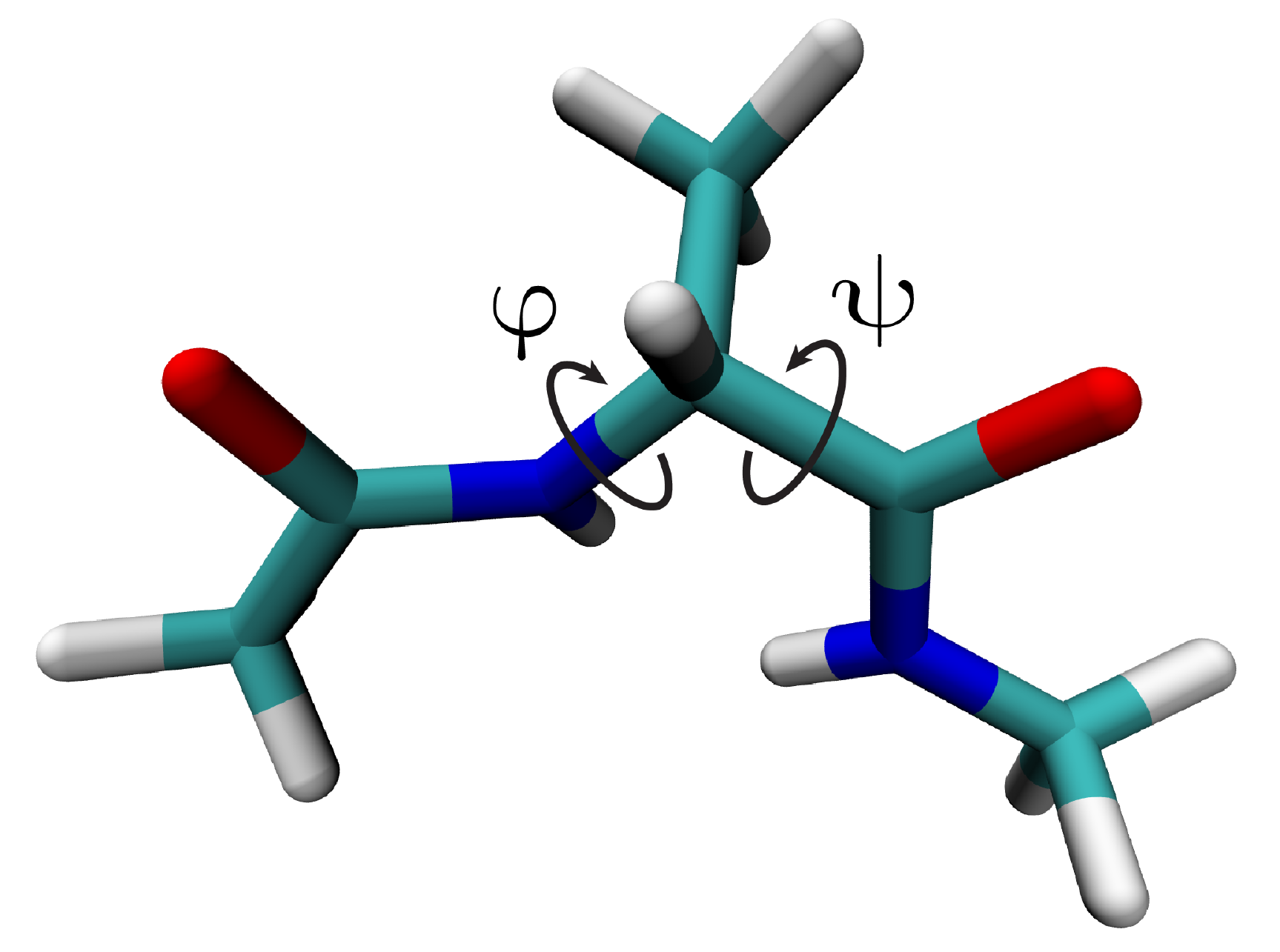}
\end{tabular}

\begin{tabular}{cc}
\footnotesize  \footnotesize \textbf{(b)} & \footnotesize \textbf{(c)}
\\
\includegraphics[width=.22\textwidth]{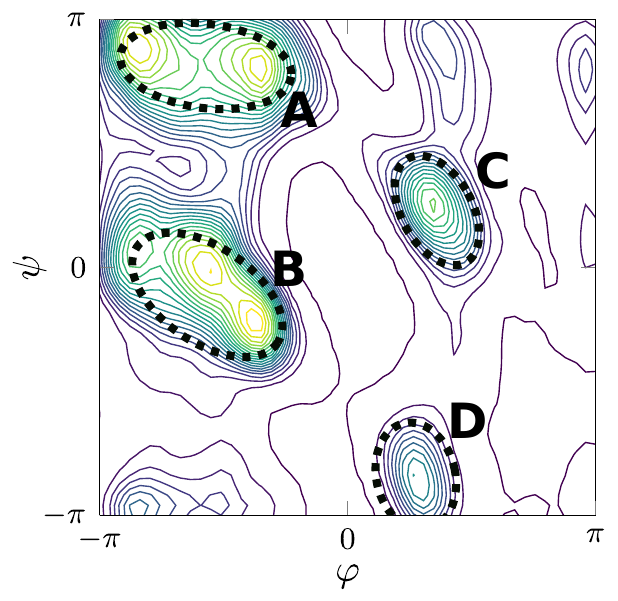}
&
\includegraphics[width=.24\textwidth]{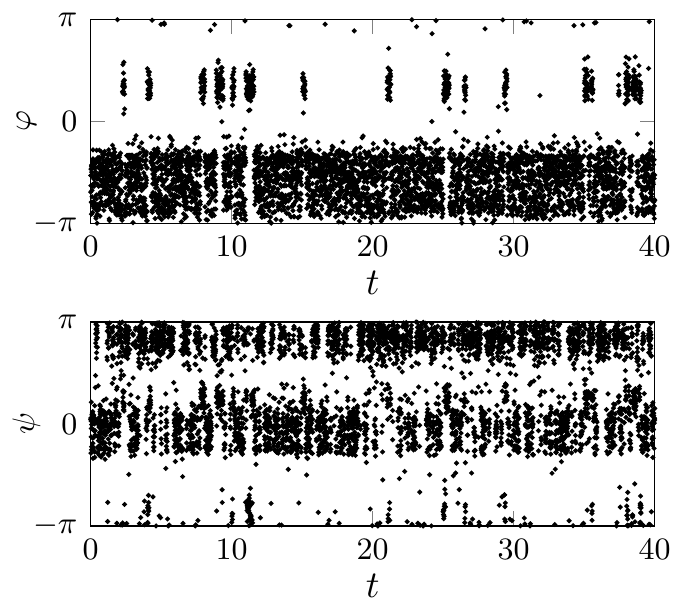}
\end{tabular}
\caption{(a)~Alanine dipeptide with its two essential dihedral angles $\varphi$ and $\psi$. (b)~Ramachandran plot of $\varphi,\psi$, revealing four metastable conformations $A,B,C,D$. (c)~Frames from a long trajectory projected onto $\varphi$ and~$\psi$. We observe the characteristic jumping pattern between the metastable sets.}
\label{fig:Dialanine}
\end{figure}

For this, we consider a single alanine dipeptide molecule in aqueous solution at temperature $400$K.
The molecule consists of 22 atoms (including hydrogen atoms), thus the full Cartesian state space $\mathbb{X}$ is 66-dimensional. 
We chose to analyze this rather small example system as it still possesses a clearly-defined timescale separation that bigger systems often lack. Furthermore, the system possesses a chemically intuitive reaction coordinate that will serve as a benchmark:
usually, two backbone dihedral angles $\varphi,\psi$ are considered responsible for the long-term kinetics of alanine dipeptide, with four configurations of these angles forming metastable states (see Figure \ref{fig:Dialanine}).
We emphasize however that this information is used only for illustration and comparison purposes and that we compute our reaction coordinate $\xi$ based on the full 66-dimensional data.

The relaxation time $t=20$ ps as well as the embedding dimension $r=2$ are assumed to be known. We will see later that $t$ indeed falls into a timescale gap. \todo{Andreas: Can we find a quick argument that this info is available in practice, or cheaply computable?}
For the dynamical data, a single $40$ ns long trajectory of the system was generated using the MD software Gromacs.
The trajectory was stripped from the solvent molecules, downsampled to step width $\tau=0.02$ ps, and its center of mass fixed at the center of the simulation box, yielding the $66$-dimensional trajectory
$$
\{x_0,\Phi^\tau x_0,\ldots, \Phi^{M\tau} x_0\}
$$
with $M=2\cdot 10^6$. Using \eqref{eq:TrajDataSets}, we generated the data sets~$\X_M,~\mathbb{Y}_M$.

We computed $2000$ Voronoi centers in the region covered by the trajectory using both the k-means- and the picking algorithm. The projection of these points onto the $(\varphi,\psi)$-plane can be seen in Figure \ref{fig:DialanineRC} (b). While this projection offers only an incomplete insight into the distribution of the full 66-dimensional center points, it indicates that the k-means algorithm again emphasizes the metastable sets, whereas Algorithm~\ref{algo:PoissonSampling} covers the total range of values more evenly. Again, for the evaluation points $\{x_1,\ldots,x_L\}$, we re-purposed the 2000 Voronoi center points.

For the embedding functions $\eta:\mathbb{R}^{66}\rightarrow \mathbb{R}^5$, linear functions with coefficients drawn uniformly randomly from $[0,1]$ were chosen just as in Section \ref{sec:BananaPot}.

\subsubsection*{Results}
\begin{figure*}
\centering

\setlength{\tabcolsep}{4pt}
\renewcommand{\arraystretch}{2.5}

\makebox[\textwidth][c]{
\begin{tabularx}{\textwidth}{cc@{\hspace{-2\tabcolsep}}c | c@{\hspace{-2\tabcolsep}}c}
&  \multicolumn{2}{c|}{\footnotesize\textbf{k-means}} & \multicolumn{2}{c}{\footnotesize\textbf{picking algorithm}} \vspace{-2pt}\\
\raisebox{12mm}[4mm][2mm]{\footnotesize\textbf{(a)}}\hspace{-10pt}
&
\parbox[c]{.23\textwidth}{
\centering
$\xi$, first component

\includegraphics[scale=0.6]{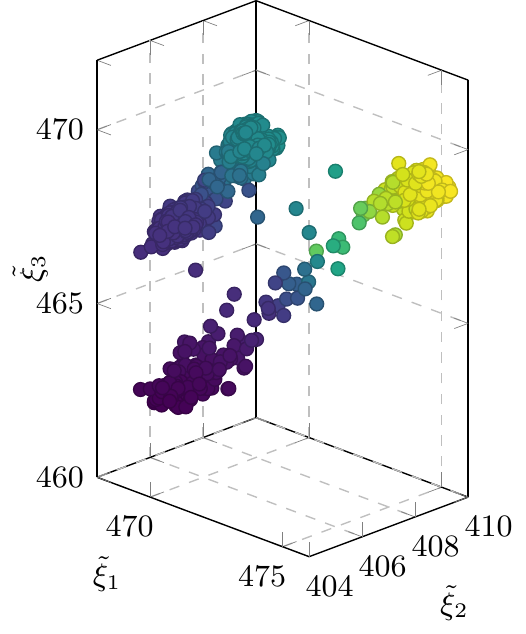}
}
&
\parbox[c]{.23\textwidth}{
\centering
$\xi$, second component

\includegraphics[scale=0.6]{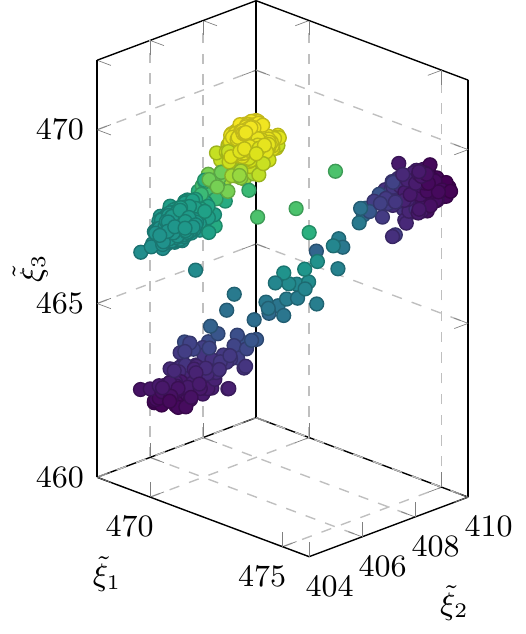} 
}
&
\parbox[c]{.23\textwidth}{
\centering
$\xi$, first component

\includegraphics[scale=0.6]{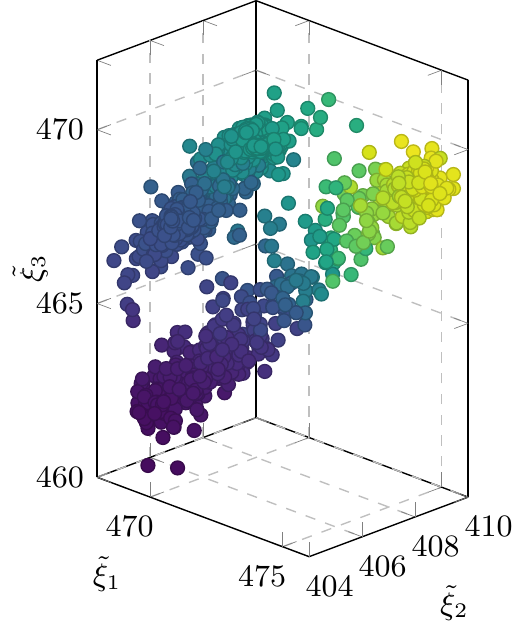}
}
&
\parbox[c]{.23\textwidth}{
\centering
$\xi$, second component

\includegraphics[scale=0.6]{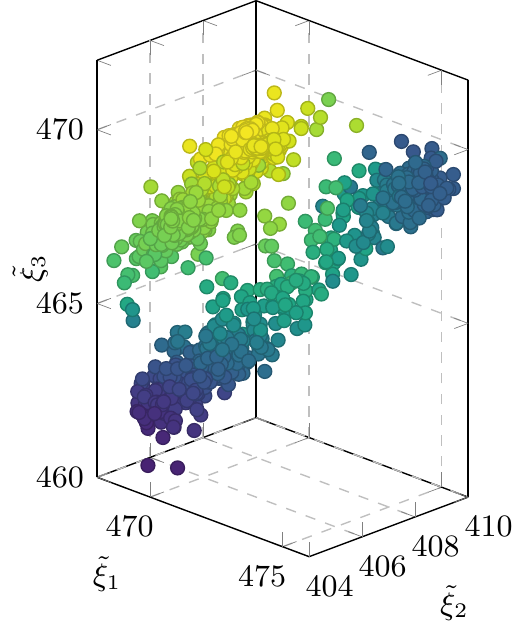}
}
\\
\raisebox{12mm}[4mm][2mm]{\footnotesize\textbf{(b)}}\hspace{-10pt}
&
\parbox[c]{.25\textwidth}{
\includegraphics[scale=0.6]{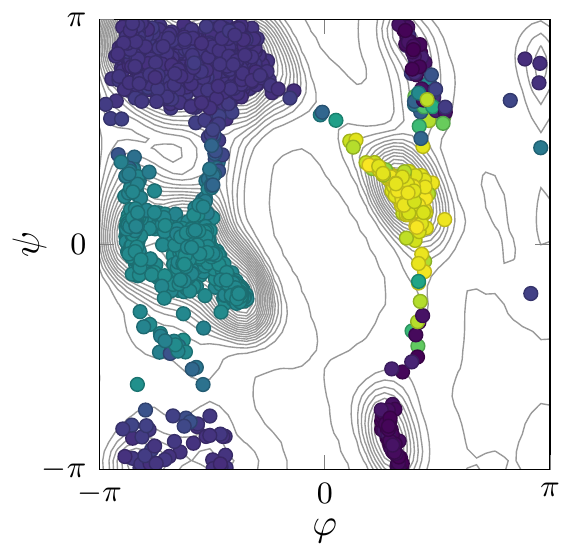}
}
&
\parbox[c]{.25\textwidth}{
\includegraphics[scale=0.6]{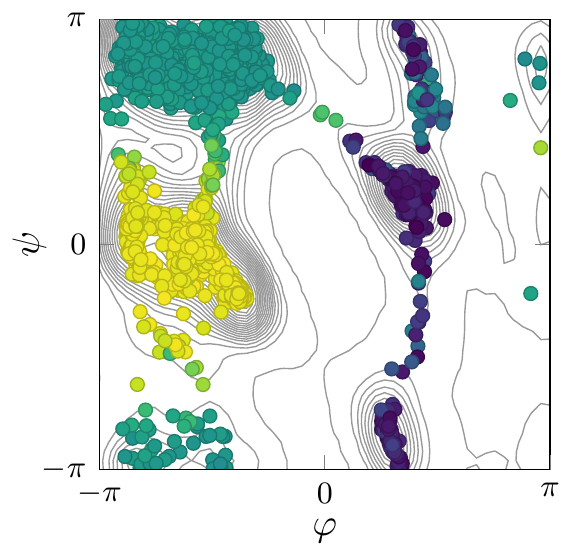}
}
&
\parbox[c]{.25\textwidth}{
\includegraphics[scale=0.6]{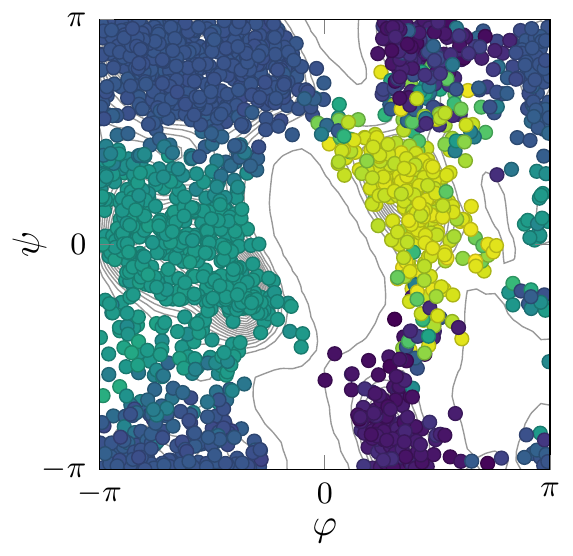}
}
&
\parbox[c]{.25\textwidth}{
\includegraphics[scale=0.6]{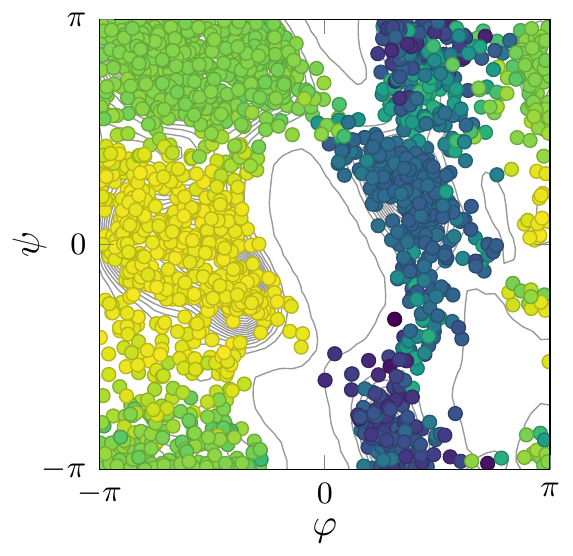}
}
\end{tabularx}
}

\caption{Alanine dipeptide reaction coordinates. (a)~The first three out of five components of the pre-reaction coordinate $\tilde{\xi}$. The output of Diffusion Maps, which is used as the final two-dimensional reaction coordinate $\xi$, are used as color map. (b)~Comparison of $\xi$ and the two dihedral angles.}
\label{fig:DialanineRC}
\end{figure*}

Figure \ref{fig:DialanineRC} visualizes the computed reaction coordinates with Voronoi center points chosen by the k-means algorithm (left) and picking algorithm (right). 

As the dimension of the transition manifold was assumed to be $r=2$, the dimension of the embedding space and thus the values $\tilde{\xi}(x_i),~i=1,\ldots,L$, is $2r+1=5$, which makes it impossible to directly visualize the embedded transition manifold.
However, plotting just the first three of the five components still offers a good insight into the structure of the embedded transition manifold, see Figure~\ref{fig:DialanineRC}~(a).
Unlike in the first example, the two-dimensional manifold structure in the embedded points is not obviously apparent. Instead, the points $\tilde{\xi}(x_i)$ appear to be mainly concentrated around four clusters, that form two connected pairs.
The Diffusion Maps algorithm still recognizes the point cloud as parametrizable by a two-dimensional coordinate and computes the parametrization i.e. our final reaction coordinate $\xi$ at the evaluation points.
Figure \ref{fig:DialanineRC} (a) and (b) show in color the two components of $\xi$ at the embedded evaluation points, and at the $(\varphi,\psi)$-projection of the evaluation points, respectively. The latter confirms that the observed four clusters correspond to the four metastable states, and the connections between the pairs of clusters corresponds to points that are located along the transition pathways. It also explains why there is seemingly no connection between the two pairs of clusters: the transition pathway connecting clusters A and C is too sparsely populated by evaluation points -- especially in the k-means case -- in order to show the connection.
Overall, we see a clear correlation between the computed reaction coordinate $\xi$ and the reference reaction coordinate $(\varphi,\psi)$.

\subsubsection*{Timescale analysis}

We again compute the implied timescales of the reduced process $\xi(\bm{X}_t)$. To yield the highest accuracy possible for the given data set, we utilize the PyEMMA software package \cite{pyemma} with its built-in methods to discretize the transfer operator, estimate its eigenvalues and compute the timescales. 

Computing the timescales of the full $66$-dimensional process with the necessary accuracy is not possible, so we cannot conduct a rigorous error analysis for this system.
Instead, we utilize the variational principle of conformation dynamics \cite{variational2013} which states that the timescales of the full process are always \emph{underestimated} by those of any projection of the process. Thus, larger dominant timescales of the projected process in general correspond to a better reaction coordinate. 
However, due to the possibility of systematic errors in approximating the projected timescales (discretization of the transfer operator, finite amount of dynamical data), this variational principle might be violated. Thus, we additionally offer a comparison to the timescales of a manually-chosen two-dimensional reaction coordinate that can generally be considered ``good'', namely the backbone dihedrals $(\varphi,\psi)$. Still, we emphasize that these timescales do not represent the ``ground truth''. The coordinate $(\varphi,\psi)$ is also not necessarily optimal in the sense of the variational principle, and thus again gives only an approximation of the full system's true dominant timescales.

Using these two error estimators, we compare our reaction coordinates $\xi$ for both the k-means and the picking algorithm to a two-dimensional TICA (time-lagged independent component analysis) projection, a dimensionality reduction method that is popular in MD analysis \cite{tika2013}.
TICA finds the directions in the data sets with maximal global autocorrelation for a specified lag time, and thus always yields \emph{linear} reaction coordinates. For this lag time $\tau=120$~ps was chosen as it maximizes the cumulative kinetic variance (95.5\%)\cite{kineticdistance}. 

The three (nontrivial) dominant timescales and their deviation from the benchmark $(\varphi,\psi)$-projection can be seen in Table \ref{tab:DialanineTimescales}. The remaining timescales $t_i,~i\geq 4$ are significantly smaller ($< 5$~ps) and are considered non-dominant and thus irrelevant.

Judging by both the variational principle and the comparison to the benchmark projection, both of our new reaction coordinates provide a measurably better approximation of the dominant timescales than the TICA reaction coordinate, though the latter remains competitive. 

\begin{table}
\vspace{0.3cm}
\centering

dominant timescales~[ps]
\vspace{0.5em}

\setlength{\tabcolsep}{0.5em}
\begin{tabular}{r | c c c }	
          & $t_1$     & $t_2$     & $t_3$      \\
    \hline\addlinespace
    \textbf{\smash{${\xi}$, k-means alg.}}     		& 194.58     & 62.50     & 41.80     \\
    \textbf{\smash{${\xi}$, picking alg.}}     		& 194.41     & 62.25     & 41.63     \\
    \textbf{\smash{TICA}}	    		& 191.78     & 61.27     & 29.84     \\
    \textbf{\smash{$(\varphi,\psi)$}}    & 194.71     & 62.93     & 41.27     \\    
    \bottomrule
\end{tabular}%
\vspace{0.5cm}
 

\caption{Dominant implied timescales of the dipeptide system under projection onto different reaction coordinates. In general, larger dominant timescales indicate better reaction coordinates. The zero-th timescale is $t_0=\infty$ in all four cases.}
\label{tab:DialanineTimescales}
\end{table}

\subsubsection*{Eigenfunction reconstruction}

As the reaction coordinate $\xi$ was constructed to fulfill Criterion $\eqref{eq:optcrit2}$, it should be possible to reconstruct the full system's dominant transfer operator eigenfunctions $v_i,~i=1,2,3$, which are functions over the $66$-dimensional state space $\X$, from the eigenfunctions $w_i$ of the projected transfer operator, i.e. functions over $\mathbb{R}^2$. As the reaction coordinates computed with the k-means and the picking algorithm variant of Algorithm \ref{algo:GalerkinRC} are qualitatively equal, we limit the investigation to the k-means reaction coordinate.

Even though the state space is 66-dimensional, the eigenfunctions of the full transfer operator can still be approximated with reasonable accuracy by a Galerkin method if an appropriate mesh-free basis is used. Luckily, we have already constructed such a basis, namely the Voronoi basis used for computing the reaction coordinates.
Thus, we are able to re-use exactly the same transition matrix $T$ and Gram matrix $S$ assembled in Algorithm \ref{algo:GalerkinRC}.
Computing the Galerkin approximation of the eigenfunctions $v_i$ then corresponds to solving a $2000\times 2000$ eigenvector problem.

On the other hand, as $\xi$ is only two-dimensional, computing the eigenfunctions $w_i$ of the projected transfer operator $\mathcal{T}^t_\xi$ is possible by a fine grid-based Galerkin method. To construct the corresponding transition matrix, the projected trajectory
$$
\big\{\xi(x_0),\xi\big(\Phi^\tau x_0\big),\ldots,\xi\big(\Phi^{M\tau} x_0\big)\big\}
$$
is used. The functions $\hat{v}_i(\cdot) := w_i\big(\xi(\cdot)\big)$ then should reconstruct the $v_i$.

\begin{figure*}
\centering

\setlength{\tabcolsep}{3pt}
\renewcommand{\arraystretch}{1}
\begin{tabular}{cccc}
\\ \vspace{-4pt}
\raisebox{34mm}[4mm][2mm]{\footnotesize \textbf{(a)}}
&
\includegraphics[scale=.9]{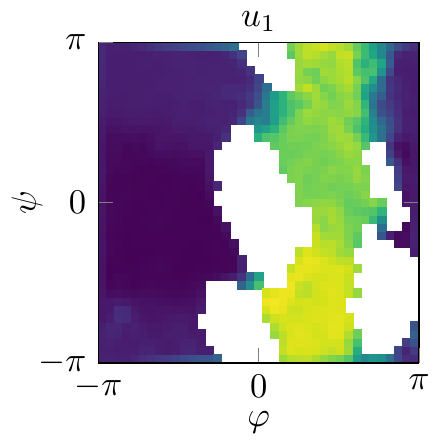}
&
\includegraphics[scale=.9]{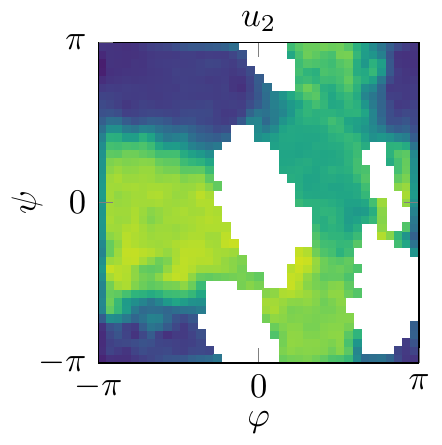}
&
\includegraphics[scale=.9]{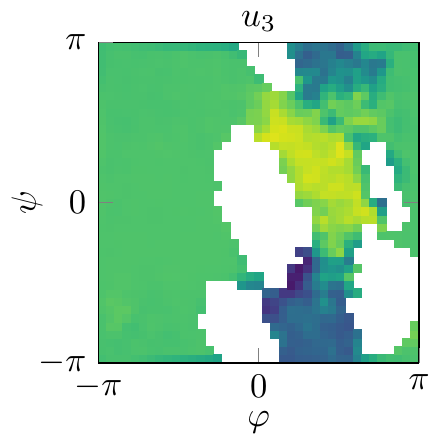}
\\ \vspace{-4pt}
\raisebox{34mm}[4mm][2mm]{\footnotesize \textbf{(b)}}
&
\includegraphics[scale=.9]{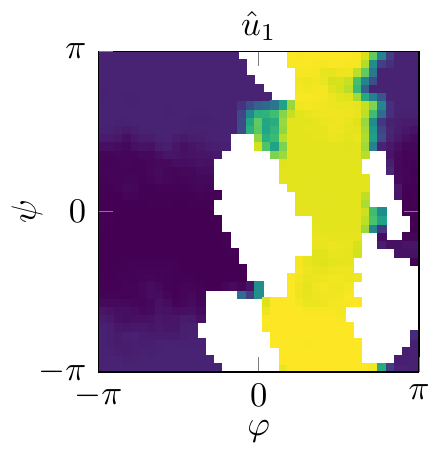}
&
\includegraphics[scale=.9]{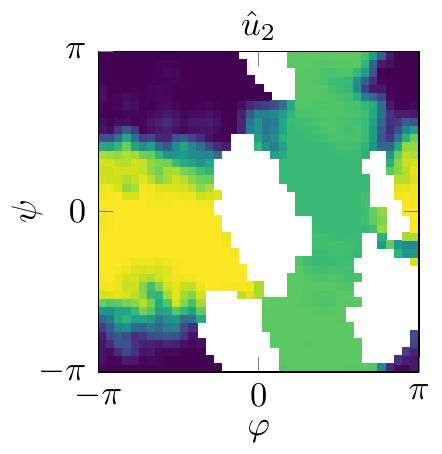}
&
\includegraphics[scale=.9]{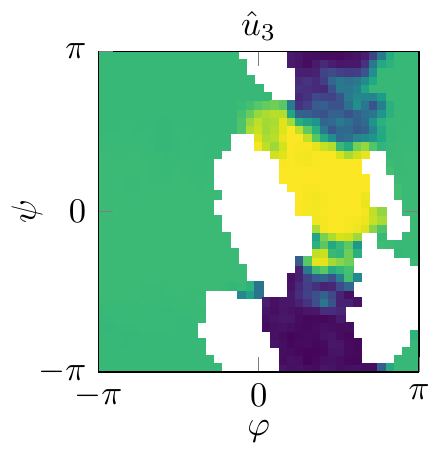}
\end{tabular}
\caption{(a)~Dominant eigenfunctions $v_i$ of the full transfer operator, projected onto the $(\varphi,\psi)$-plane. (b)~Reconstructed eigenfunctions $\hat{v}_i$, projected onto the $(\varphi,\psi)$-plane. Regions that do not contain trajectory data are left white.}
\label{fig:DialanineEF}
\end{figure*}

Of course, being functions over the 66-dimensional state space, the $v_i$ and $\hat{v}_i$ are difficult to visualize. We thus again project them onto the $(\varphi,\psi)$-plane using a simple interpolation procedure.
The result can be seen in Figure~\ref{fig:DialanineEF}. We observe excellent qualitative aggreement between the full and the reconstructed eigenfunctions, or at least their $(\varphi,\psi)$-projections.

\begin{remark*}
This last section has again shown the close relationship between the transfer operator eigenfunctions and the newly-defined reaction coordinates, both in their expressive power as well as the data required to compute them.
In this concrete example, even the computational effort is identical, as both the computation of the full transfer operator eigenfunctions as well as the application of the Diffusion Maps algorithm to the embedded evaluation points requires the solution of a $2000\times 2000$ eigenproblem. 
Therefore, our proposed numerical method is not necessarily computationally advantageous over directly computing the eigenfunctions.

However, we want to stress again that the newly-defined transition manifold-based reaction coordinates are advantageous on a conceptual level.
Firstly, they obey a rigorous optimality criterion and thus are guaranteed to preserve the system's slowest timescales.
Secondly, they are interpretable in the context of transition pathways, as detailed in Section \ref{sec:ConnectionTPT}. For alanine dipeptide, the computed two-dimensional reaction coordinate $\xi$ is directly interpretable as a transformation of the two principal dihedral angles, whereas using the three dominant eigenfunctions as reaction coordinates would yield a three-dimensional reaction coordinate that is redundant in describing the molecule's internal slow dynamics.

\end{remark*}

\subsection{Conformational analysis of NTL9}

Finally, we demonstrate the applicability of our method to a realistic high-dimensional system. 
For this, we analyze a 1.11 ms long molecular trajectory of the fast-folding protein NTL9, generated on the Anton supercomputer\cite{Lindorff-Larsen11}.
Instead of cartesian coordinates, we use the amino acid chain's \emph{contact map}, i.e., the matrix containing the pairwise distances of the residues, as coordinates for the further analysis. This eliminates the need to remove the global translational and rotational motion from the trajectory. As the protein consists of 40 residues, this results in a 1600-dimensional state space (although it could be reduced due to symmetry of the contact map matrix).

As we are interested in the forming of secondary structures such as $\alpha$ helices and $\beta$ sheets, we choose a lag time 1-2 orders of magnitude faster than those processes, $\tau=10$\,ns.
To generate the data set $\mathbb{X}_M$, $M=1.11\cdot 10^6$ frames were uniformly subsampled from the trajectory. $\mathbb{Y}_M$ was generated the same way, only with a lag time of $\tau$. 
From $\mathbb{X}_M$, we drew $L=5550$ Voronoi center points $x_i$ using the picking algorithm.

For the expected transition manifold dimension $r$, and the corresponding number of embedding functions $2r+1$, we used a simplified version of the iterative procedure proposed at the end of Section \ref{sec:TMembedding}: Start with a low value ($r=1$ in this example), and see if useful structure can be identified in the embedded transition manifold. If not, increase $r$ and repeat the embedding procedure. 

\subsubsection*{Results.}

\begin{figure*}
\centering

\begin{tabular}{c c}
\raisebox{68mm}[4mm][2mm]{\footnotesize \textbf{(a)}}
&
\includegraphics[scale=1]{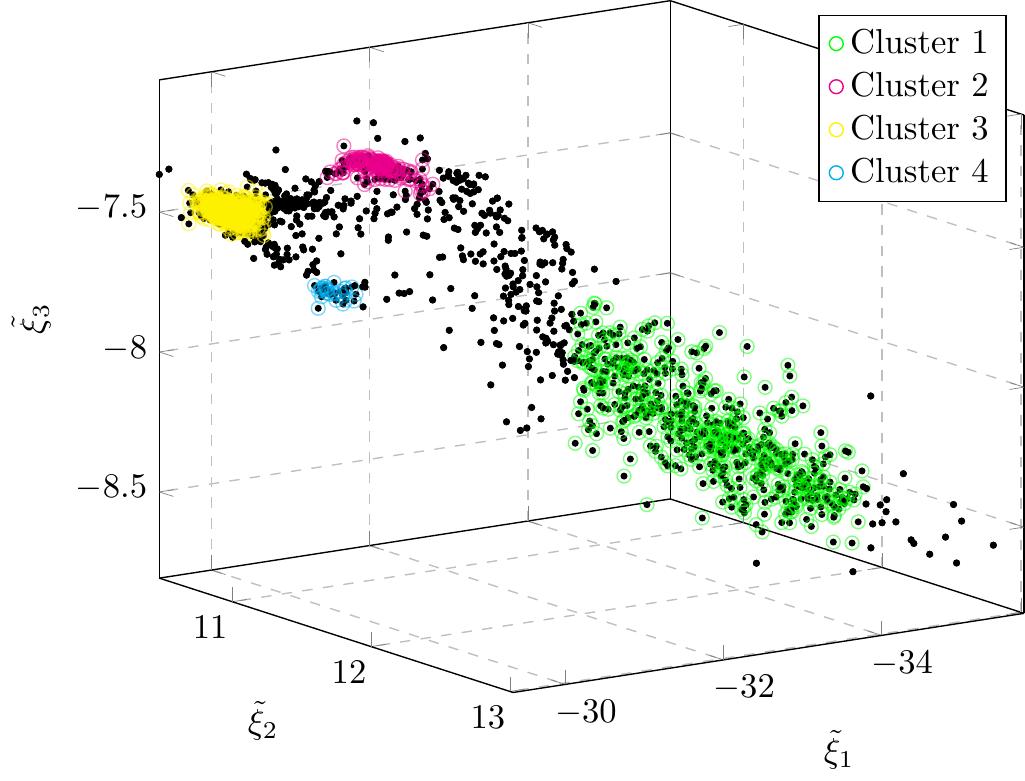}
\end{tabular}

\begin{tabular}{c c c c c}
\raisebox{32mm}[4mm][2mm]{\footnotesize \textbf{(b)}}
&
\includegraphics[scale=1]{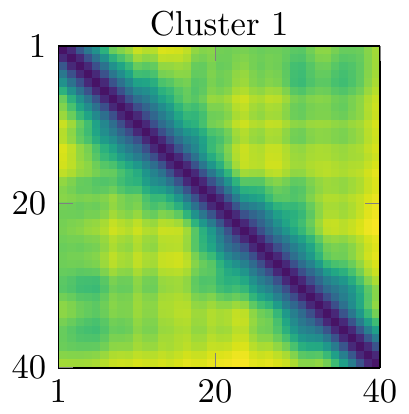}
&
\includegraphics[scale=1]{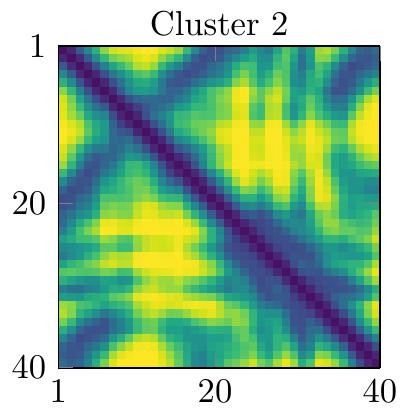}
&
\includegraphics[scale=1]{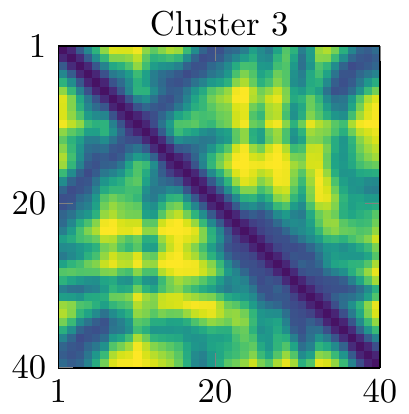}
&
\includegraphics[scale=1]{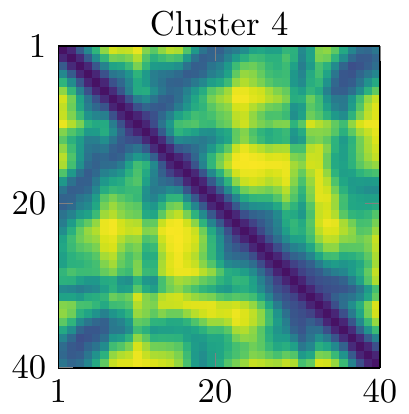}
\\
\raisebox{24mm}[4mm][2mm]{\footnotesize \textbf{(c)}}
&
\includegraphics[width=.23\textwidth]{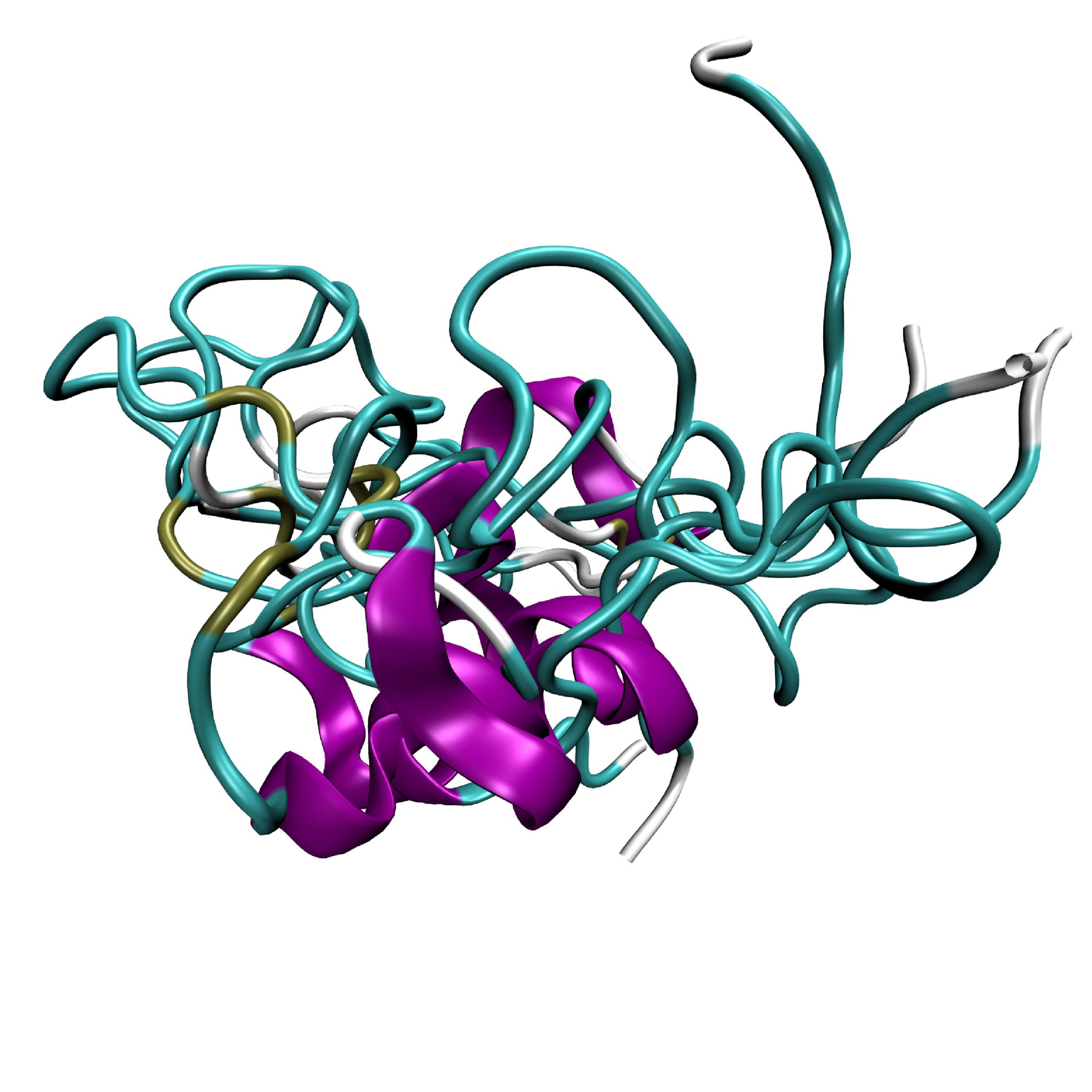}
&
\includegraphics[width=.23\textwidth]{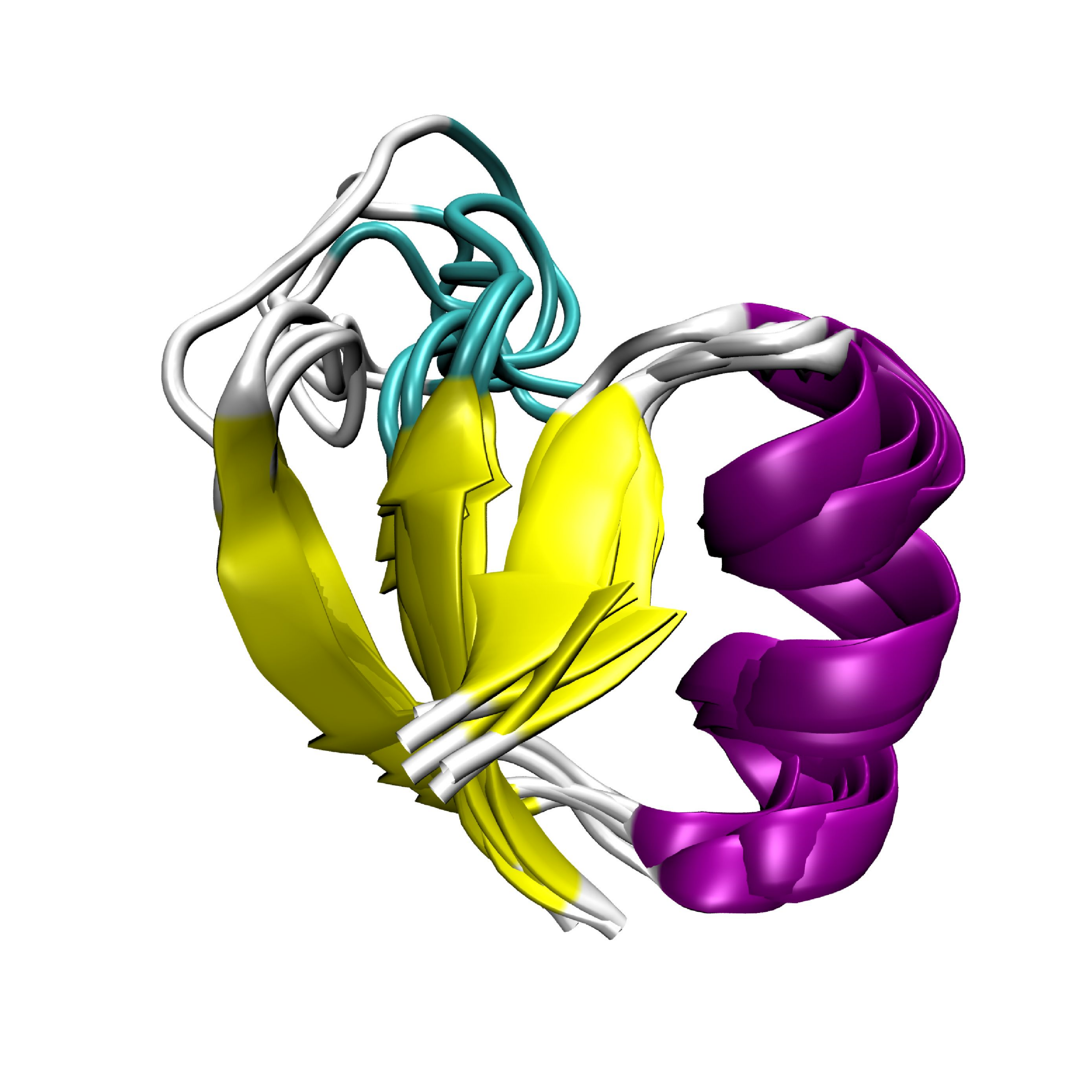}
&
\includegraphics[width=.23\textwidth]{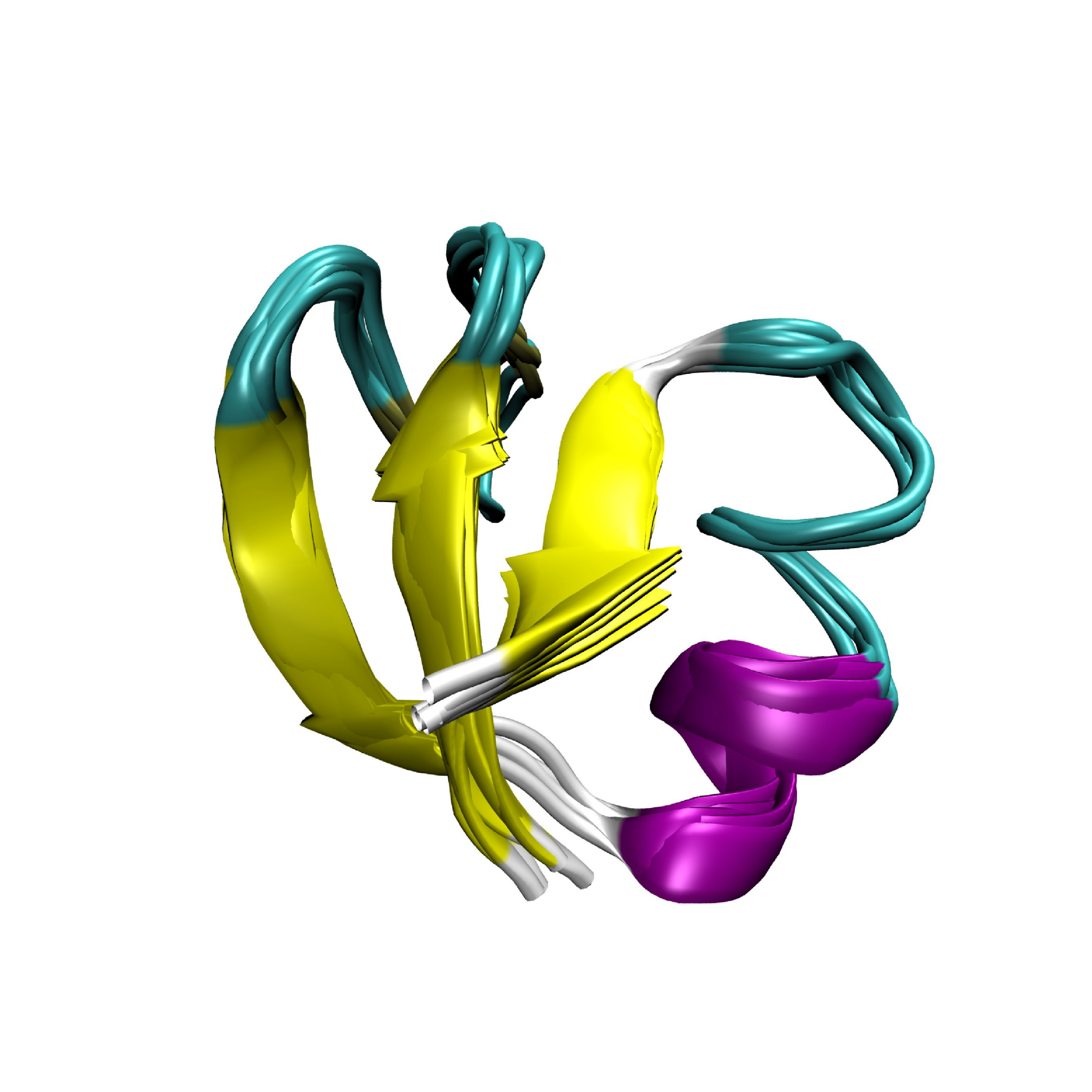}
&
\includegraphics[width=.23\textwidth]{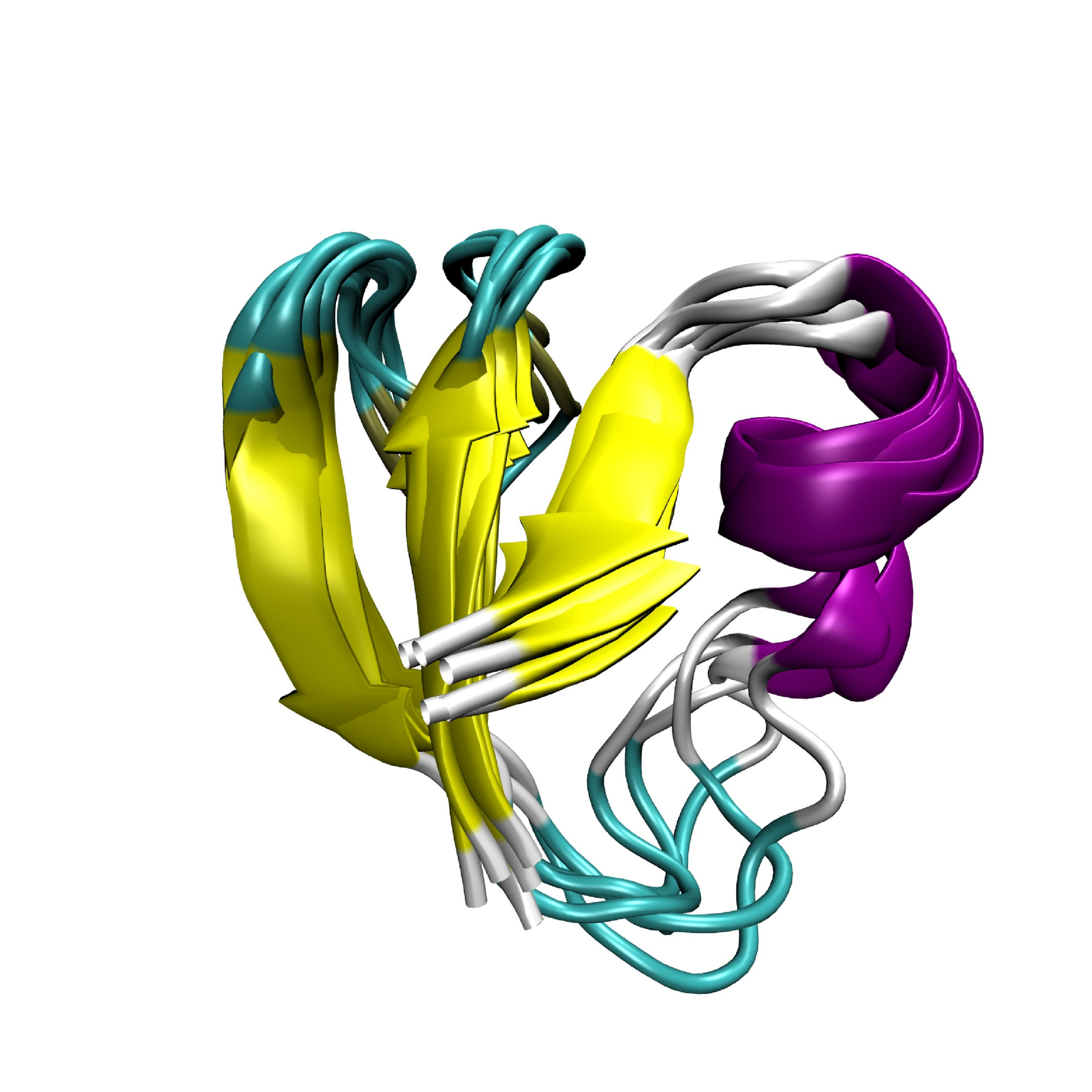}
\\
&
Unfolded (8.5\%)
&
Folded 1 (2.9\%)
&
Folded 2 (79.4\%)
&
Folded 3 (0.5\%)
\end{tabular}

\caption{NTL9 transition manifold. (a) Embedded points $\tilde{\xi}(x_i)$, representing the embedded transition manifold embedded into $\mathbb{R}^3$. Four clusters are visible. (b) Average contact maps of all points in the clusters. (c) Overlay  of the secondary structures of five randomly-selected points per cluster.}
\label{fig:NTL9_clusters}

\end{figure*}

Quite surprisingly, the transition manifold in this case already reveals its structure under an embedding into $\mathbb{R}^3$.
In the embedded Voronoi center points $\tilde{\xi}(x_i)$, four clusters are clearly visible (Figure \ref{fig:NTL9_clusters} (a)). The clusters are robust under the choice of the embedding functions. 

For simplicity, i.e., in order to avoid the parameter tuning of an automated clustering algorithm, we assigned the points to the clusters manually. Their average contact maps and secondary structure are shown in Figure \ref{fig:NTL9_clusters}~(b) and~(c). 

Interpreting the four clusters as conformations, our results are to a large degree consistent with those of Mardt et al. \cite{Mardt2018}, who performed analysis on the same dataset using deep learning methods. Our conformations ``Unfolded'', ``Folded 1'' and ``Folded 2'' correspond very well to the main conformations identified by their algorithm. Note that ``Unfolded'' is not a conformation in the chemical sense, but rather a loose collection of various unfolded configurations. The populations of the conformations (percentages in Figure \ref{fig:NTL9_clusters} (c)) are also comparable to those in Ref.~\onlinecite{Mardt2018}. Our slightly lower values can be explained by the difference in how the populations are calculated. However, our conformation ``Folded 3'' does not appear in their analysis. While its population is quite low, its structure subtly yet distinctively differs from the other conformations, so we do not consider its existence a statistical artifact. Furthermore, we were not able to find the finer sub-structures of the ``Unfolded'' conformation that were identified in Ref.~\onlinecite{Mardt2018}. 

\begin{figure*}
\centering
\begin{minipage}{.5\textwidth}
\begin{tabular}{c c}
\raisebox{46mm}[4mm][2mm]{\footnotesize \textbf{(a)}}
&
\includegraphics[scale=1]{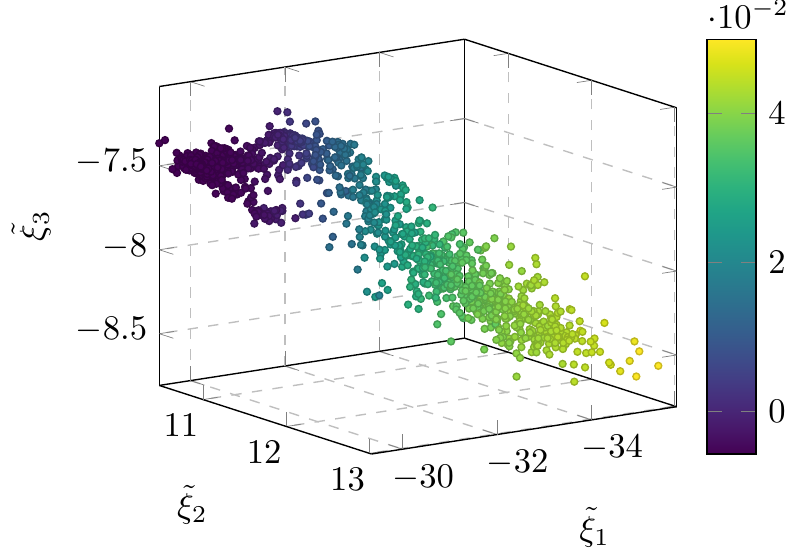}
\end{tabular}
\end{minipage}
\hspace{1em}
\begin{minipage}{.38\textwidth}
\begin{tabular}{c c}
\raisebox{42mm}[4mm][2mm]{\footnotesize \textbf{(b)}}
&
\includegraphics[scale=1]{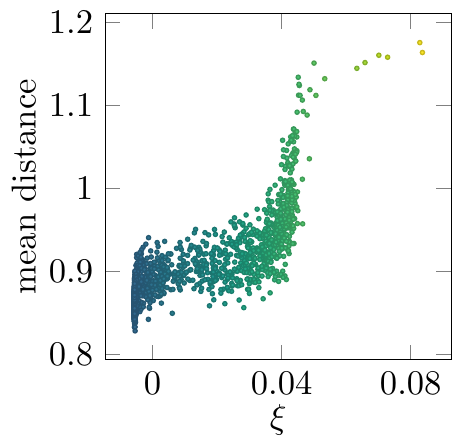}
\end{tabular}
\end{minipage}

\caption{Diffusion map analysis of the embedded transition manifold. (a) First diffusion map coordinate $\xi$ in the embedded Voronoi center points. (b) Comparison of $\xi$ and the mean residual distance shows a nonlinear, yet monotonous correlation.}
\label{NTL9:DiffMaps}
\end{figure*}

Let $\xi(x_i)$ denote the first diffusion maps coordinate on the embedded points, which indicates the direction of largest variance.
We see a strong correlation between $\xi$ and the mean inter-residuum distance, i.e., the average of all entries of the contact map matrix (Figure \ref{NTL9:DiffMaps}). Thus, $\xi$ describes the "degree of foldedness" of the protein, which can be considered a reasonable one-dimensional reaction coordinate of this system.
However, unlike in the Dialanine example, the second and higher diffusion map coordinates here did not correspond to some easily-interpretable physical property that finer resolves the transitions between the identified conformations, and instead seemed to consist only of higher modes of the first diffusion map coordinate.

\begin{remark*}


Although the results are already very encouraging and show the potential usefulness of the method for very high-dimensional systems, the setup can be refined in a number of ways. 
Most importantly, instead of the simple Voronoi cell-based Galerkin method, specialized ansatz spaces such as meshfree basis functions with global support, might be able to better approximate the reaction coordinate, in particular in the undersampled transition regions.
We are planning to explore these and other refinements of the method as well as its application to further high-dimensional molecular systems in an upcoming work.
\end{remark*}

\section{Conclusion}
\label{sec:conclusions}

In this paper, we reviewed a novel framework for the characterization and computation of optimal reaction coordinates, originally introduced in Ref.~\onlinecite{Bittracher2017} and presented efficient algorithms for the computational identification of such reaction coordinates that allow for direct application to real-world molecular systems. Moreover, we found that the new framework agrees with the TPT characterization of good reaction coordinates in classical metastable systems, but offers more rigorous criteria that are applicable to much broader classes of multiscale systems with and without timescale gap.

In particular, we introduced a discretization approach to the data-driven computation of reaction coordinates that fulfill these rigorous criteria. This approach is usable whenever a transition matrix between the discretization elements can be computed from available simulation data, e.g., if the data represent a long (equilibrated) trajectory so that the entire machinery invented for building MSMs can now be utilized for the computation of reliable reaction coordinates with provable approximation quality.

As a demonstration, we provided two algorithms to construct a meshfree basis of Voronoi ansatz functions directly from the data. Both algorithms are highly scalable and readily available, making this method straightforward to apply for practitioners who have existing simulation data on their hard drives. We showed that in molecular systems of medium size, the resulting approximation error in the dominant time scales is competitive with state of the art dimension reduction techniques. This demonstrates that the reaction coordinates we compute here can be used to build efficient coarse grained models. Of course the computed reaction coordinates themselves are also of independent value. In the case of the alanine dipeptide we showed that our computed reaction coordinates and the dihedral angles, which are typically used as reaction coordinates for this system, produce a similar portrait of the system when viewed in this reduced space.

As a next step, we plan to apply our method to higher-dimensional systems with a priori unknown reaction coordinates, using specialized Galerkin ansatz spaces borrowed from Markov State Model theory. For these systems however, the requirement to have simulation data that samples the stationary density is of course somewhat strict. We thus also plan to work on relaxing this requirement to a ``local'' version, i.e., we will work with samples that are equilibrated only in some smaller region of the state space, in order to compute reaction coordinates in that region.

\section*{Acknowledgements}

This research has been funded by Deutsche Forschungsgemeinschaft (DFG) through grant CRC 1114 ``Scaling Cascades in Complex Systems''.

\begin{appendix}

\section{Spectral decomposition of the transition density function}
\label{ap:TransdensityDecomp}

We derive the exact form of the coefficients $d_i(x,t)$ in the decomposition of the transition density function $p^t(x,\cdot)$, 
$$
p^t(x,\cdot) = \sum_{i=0}^\infty d_i(x,t) v_i(\cdot).
$$
Recall that the stochastic process was assumed to be \emph{reversible}, which formally equates to the transition density function and the stationary density $\rho$ fulfilling the \emph{detailed balance condition}
$$
p^t(x,y)\rho(x) = p^t(y,x) \rho(y) \quad \text{for all } x,y\in\X~.
$$

With that, it is easy to see that the transfer operator $\mathcal{P}^t$, defined in \eqref{eq:transferoperator}, is self-adjoint with respect to the weighted inner product
$$
\langle f,g\rangle_{\rho^{-1}} := \int_\X f(x) g(x) \frac{1}{\rho(x)}~dx:
$$
\begin{align*}
\big\langle \mathcal{P}^tf,g\big\rangle_{\rho^{-1}} &= \int_\X \mathcal{P}^tf(x)g(x)\frac{1}{\rho(x)}~dx \\
&=\int_\X \Big(\int_\X f(y) p^t(y,x)~dy\Big) g(x)\frac{1}{\rho(x)}~dx \\
&=\int_\X f(y) \Big(\int_\X g(x) p^t(y,x)\frac{1}{\rho(x)}~dx\Big)~dy \\
&\overset{(*)}{=}\int_\X f(y)  \Big(\int_\X g(x) p^t(x,y)~dx\Big)\frac{1}{\rho(y)}~dy \\
&=\big\langle f,\mathcal{P}^tg\rangle_{\rho^{-1}}~,
\end{align*}
where in $(*)$ the detailed balance condition was used.
Thus, the eigenfunctions $v_i$ of $\mathcal{P}^t$ form an orthogonal basis of the associated inner product space.

We assume from now on that the function $p^t(x,\cdot)$ lies in (or can be approximated with sufficient accuracy) in this space. Then we have
\begin{align*}
p^t(x,y) &= \sum_{i=1}^\infty \big\langle p^t(x,\cdot),v_i\big\rangle_{\rho^{-1}}~ v_i(y)\\
\end{align*}
Now, $p^t(x,\cdot)$ can be seen as the time-$t$ evolution of the Dirac density $\delta_x$ under the dynamics, thus
\begin{align*}
\big\langle p^t(x,\cdot),v_i \big\rangle_{\rho^{-1}} &= \big\langle \mathcal{P}^t\delta_x,v_i \big\rangle_{\rho^{-1}}
\intertext{Using the self-adjointness of $\mathcal{P}^t$, we get}
&= \big\langle \delta_x, \mathcal{P}^tv_i \big\rangle_{\rho^{-1}}
\intertext{As $v_i$ is an eigenfunction of $\mathcal{P}^t$ to eigenvalue $\lambda_i^t$, this is}
&= \big\langle \delta_x, v_i \big\rangle_{\rho^{-1}} \lambda_i^t.
\end{align*}
Finally, taking the inner product with $\delta_x$ is equivalent to a point evaluation at $x$:
\begin{align*}
\big\langle \delta_x, v_i \big\rangle_{\rho^{-1}} &= \int_\X \delta_x(y) v_i(y) \frac{1}{\rho(y)}~dy = \frac{v_i(x)}{\rho(x)}~.
\end{align*}
The overall decomposition thus reads
$$
p^t(x,\cdot) = \sum_{i=1}^\infty \lambda_i^t\frac{v_i(x)}{\rho(x)} v_i(\cdot)~.
$$

\section{Derivation of the Galerkin-approximated reaction coordinate}
\label{ap:GalerkinRCFormula}

Let $\mathcal{V}_N$ be a final-dimensional function space spanned by the basis $\{\varphi_1,\ldots,\varphi_N\}$. The Galerkin projection of $\tilde{\xi}$ onto $\mathcal{V}_N$ with respect to the inner product $\langle\cdot,\cdot\rangle_\rho$
is defined as
\begin{equation}\label{eq:Galerkin_xi_raw}
\tilde{\xi}_N:= \sum_{k,j=1}^N (S^{-1})_{kj} \langle \varphi_k,\tilde{\xi}\rangle_\rho \varphi_j~,
\end{equation}
with the nonnegative, symmetric Gram matrix
$$
S_{kj} = \langle \varphi_k,\varphi_j\rangle_\rho~.
$$
The Galerkin projection of the observable $\eta$ is analogously defined, and with the factors
$$
c_j  := \sum_{k=1}^N (S^{-1})_{kj} \langle \varphi_k,\eta\rangle_\rho
$$
can be written as 
$$
\eta_N:= \sum_{j=1}^N c_j \varphi_j~.
$$
We assume that the Galerkin ansatz space is suitable to approximate $\eta$, i.e. that $\|\eta - \eta_N\|_\rho$ is small, where $\|\cdot\|_\rho$ is the norm induced by $\langle\cdot,\cdot \rangle_\rho$.

Recall that $\tilde{\xi}$ can also be written as the Koopman operator applied to $\eta$: $\tilde{\xi}(x) =\mathcal{K}^t\eta(x)$.
With this, the scalar product in \eqref{eq:Galerkin_xi_raw} can be estimated as follows:
\begin{equation}
\begin{aligned}
\label{eq:GalerkinCoeffs}
\langle \varphi_k,\tilde{\xi}\rangle_\rho &= \langle \varphi_k,\mathcal{K}^t\eta\rangle_\rho \\
&\overset{(*)}{\approx} \langle \varphi_k,\mathcal{K}^t\eta_N\rangle_\rho =\sum_{l=1}^N\langle \varphi_k,\mathcal{K}^t\varphi_l\rangle_\rho c_l~.
\end{aligned}
\end{equation}
For $(*)$ it was used that $\|\mathcal{K}^t\|_\rho=1$, i.e. $\mathcal{K}^t$ does not amplify the approximation error of $\eta_N$. Define the $N\times N$ \emph{transition matrix} by
$$
T_{kl} := \langle \varphi_k,\mathcal{K}^t\varphi_l\rangle_\rho~.
$$
Then the Galerkin approximation \eqref{eq:Galerkin_xi_raw} becomes
$$
\tilde{\xi}_N(x) \approx \sum_{k,j=1}^N \varphi_j(x) (S^{-1})_{kj} \sum_{l=1}^NT_{kl}c_l~.
$$

\section{Error measure for the Voronoi center points}
\label{ap:VoronoiCenters}

We show that different choices of the error measure for approximating $\tilde{\xi}$ by its Voronoi Galerkin projection $\tilde{\xi}_N$ leads to different optimal strategies in choosing the Galerkin center points.

\paragraph{Minimizing the $L^2$ error.} 

Assume that by choosing the Voronoi center points $\{e_1,\ldots,e_N\}$ that define $\tilde{\xi}_N$, we want to minimize the error
\begin{equation}
\|\tilde{\xi} - \tilde{\xi}_N\|_\rho \overset{!}{=} \min_{\{e_1,\ldots,e_N\}\subset \mathbb{X}}~.
\end{equation}
The difficulty is that neither $\tilde{\xi}$ nor $\tilde{\xi}_N$ are known in advance.
We thus construct a Monte Carlo estimator of $\|\tilde{\xi} - \tilde{\xi}_N\|_\rho $ based on the sampled data $\mathbb{X}_M = \{x_1, \ldots, x_M\}$. Here 
\begin{equation}
\label{eq:cellmean}
\bar\xi_{A_k} = |A_k|^{-1}\sum_{x_i\in A_k} \tilde\xi(x_i)
\end{equation}
is the mean of $\tilde\xi$ in cell $A_k$. First, since $A_1,\ldots, A_N$ partition $\mathbb{X}$,
\[
\|\tilde{\xi} - \tilde{\xi}_N\|_\rho = \sum_{k=1}^N  \int_{A_k} \left(\tilde\xi(x) - \tilde\xi_N(x)\right)^2 \rho(x)~ dx.
\]
The integral can be approximated by a Monte Carlo sum over the $M$ $\rho$-distributed samples $x_i$:
\begin{align*}
\int_{A_k} & \left(\tilde\xi(x) - \tilde\xi_N(x)\right)^2 \rho(x)~dx\\
&\approx \frac{1}{M}\sum_{x_i \in A_k} \|\tilde \xi(x_i) - \tilde\xi_N(x_i)\|^2~,
\end{align*}
where $\|\cdot\|$ is the Euclidean norm in $\mathbb{R}^{2r+1}$.
Finally, since $\tilde\xi_N = \sum_k \frac{\langle \mathbbm{1}_{A_k}, \tilde \xi\rangle_\rho}{\langle \mathbbm{1}_{A_k}, \mathbbm{1}\rangle_\rho} \mathbbm{1}_{A_k}$, we may approximate $\tilde\xi_N(x)$ for $x\in A_k$ with another Monte Carlo sum:
\[
\tilde\xi_N(x) = \frac{\langle \mathbbm{1}_{A_k}, \tilde \xi\rangle_\rho}{\langle \mathbbm{1}_{A_k}, \mathbbm{1}\rangle_\rho} \approx \frac{1}{|A_k|}\sum_{x_i \in A_k} \tilde\xi(x_i) = \bar\xi_{A_k}.
\]
Combining everything gives
\begin{align*}
\|\tilde{\xi} - \tilde{\xi}_N\|_\rho &\; \approx M^{-1}\sum_{k=1}^N \sum_{x_i \in A_k} \|\tilde \xi(x_i) - \bar\xi_{A_k}\|^2_{\R^{2k+1}} \\
&\; =: S_\xi(A_1, \ldots, A_N)~.
\end{align*}

$S_\xi(A_1, \ldots, A_N)$ can be recognized as the the objective function of $k$-means clustering in the image space of the reaction coordinate $\tilde \xi$. To minimize this objective function directly one would have to know $\tilde \xi$.
If we however additionally assume that $\tilde \xi$ is Lipschitz continuous with Lipschitz constant $L$, then
\[
S_\xi(A_1, \ldots, A_N) \leq M^{-1} L^2 \sum_{k=1}^N \sum_{x_i \in A_k} \| x_i - e_k\|^2
\]
where $e_k$ is such that $\tilde \xi(e_k) = \bar \xi_{A_k}$ and $\|\cdot\|$ now is the Euclidean norm in $\mathbb{R}^n$. Minimizing this upper bound is now achieved by $k$-means clustering the data set $\X_M$ in the original state space.

\paragraph{Minimizing the uniform error.} 
Assume that we now want to minimize the uniform error
$$
\|\tilde{\xi} - \tilde{\xi}_N\|_\infty \overset{!}{=} \min_{\{e_1,\ldots,e_N\}\subset \mathbb{X}}~.
$$

Assume again that $\tilde{\xi}$ is Lipschitz continuous with Lipschitz constant $L$.
Evidently, we have
\[
\| \tilde\xi - \tilde\xi_N\|_\infty = \max_{k=1\ldots N} \sup_{x\in A_k} \| \tilde \xi(x) - \bar\xi_{A_k}\|
\]
with $\bar\xi_{A_k}$ as defined in \eqref{eq:cellmean}. Let now $e_k\in A_k$ be such that $\tilde\xi(e_k) = \bar\xi_{A_k}$ (such an $e_k$ exists by continuity of $\tilde\xi$). Then, with $\|\cdot\|_{\infty,A_k}$ denoting the uniform norm in $A_k$,
\begin{align*}
\| \tilde\xi - \tilde\xi_N\|_{\infty,A_k}  &= \sup_{x\in A_k} \| \tilde \xi(x) - \tilde\xi(e_k)\| \\
&\leq L \sup_{x\in A_k} \| x - e_k\| \\
&\leq L\: \mbox{diam}(A_k),
\end{align*}
where $\mbox{diam}(A_k)$ is the diameter of the Voronoi cell $A_k$. Since $A_1,\ldots, A_N$ partition $\mathbb{X}$, we have
\begin{align*}
\| \tilde\xi - \tilde\xi_N\|_\infty &= \max_{k=1\ldots N} \| \tilde\xi - \tilde\xi_N\|_{\infty,A_k} \\
&\leq  L\max_{k=1\ldots N}\mbox{diam}(A_k).
\end{align*}
Minimizing this upper bound then means looking for Voronoi centers such that the diameter of the largest Voronoi cell is minimized. Since the number of Voronoi cells and the volume of the set $\mathbb{X}_M$ to be covered are fixed, the minimum is achieved if the centers cover $\mathbb{X}_M$ evenly, such that the Voronoi cells all have similar diameters. Therefore, we may alternatively maximize the diameter of the smallest Voronoi cell, which is bounded from below by the minimal internal point distance:
\[
\min_{i=1\ldots N} \mbox{diam}(A_i) \geq \min_{\stackrel{i,j=1,\ldots,N}{i\neq j}} \|e_i-e_j\|.
\]
The inequality holds because $\min\|e_i - e_j\|$ is twice the distance from $e_i$ to that face of $A_i$ which is closest to $e_i$, while the diameter of $A_i$ is by definition larger. Maximizing the lower bound then leads to the objective function of maximal minimal internal point distance:
$$
E = \argmax_{\{e_1,\ldots,e_N\}\subset \mathbb{X}_M} \min_{\stackrel{i,j=1,\ldots,N}{i\neq j}} \|e_i-e_j\|~.
$$

\end{appendix}

\bibliographystyle{apsrev4-1}
\bibliography{References}

\end{document}